\documentclass[12pt,letterpaper]{article}

\usepackage{appendix}
\usepackage{multirow}

\usepackage[flushleft]{threeparttable}

\usepackage{pdflscape}

\usepackage{subcaption}

\usepackage{hyphenat}

\usepackage{bm} 

\usepackage{setspace} 
\usepackage{footmisc}
\usepackage{amsfonts}
\usepackage{amsmath}
\usepackage{theorem}

\usepackage[top=2.5cm, bottom=2.5cm, left=2.5cm, right=2.5cm]{geometry}

\usepackage{setspace}
\usepackage{footmisc}

\usepackage[natbibapa]{apacite}
\usepackage{amssymb}
\setcitestyle{aysep={}}

\usepackage{mathtools}
\usepackage{amsmath}
\usepackage[latin2]{inputenc}
\usepackage{graphicx}
\usepackage{subcaption}
\usepackage{multirow}
\usepackage{esvect}
\usepackage{afterpage}

\usepackage{lipsum}
\usepackage{booktabs,array,dcolumn}

\usepackage{titling}

\usepackage{titling}
\usepackage{blindtext}
\usepackage{hyperref}

\begin{document}

\title{\textbf{\large What do surveys say about the trend in inequality and the applicability of two table-transformation methods?}} 

\author{Anna Naszodi}

\date{}

	\begin{titlepage}
		\date{}
		
		\title{\textbf{\large What do surveys say about the trend in inequality and the applicability of two table-transformation methods?}}\thanks{
			{\textit{Email}: naszodi@idil.li. \textit{OrcidID}: 0000-0002-3880-6566. 
				 Anna Naszodi is the founding director of the  \href{https://idil.li}{International Demographic Inequality Lab} 	(see: \url{https://idil.li}). 
				 Also, she is an honorary member of the \href{https://kti.krtk.hu/en/researchers/naszodi-anna/}{KRTK-KTI} and a senior researcher affiliated with the Central Bank of Hungary. 
				She started to work on this paper while being employed by the Joint Research Centre of the European Commission.\\ 
				\textit{Disclaimer}: The views expressed in this paper are those of the author and do not necessarily reflect the official views of the European Commission and the Central Bank of Hungary.}}
		
		\author{ANNA NASZODI}

		\maketitle
		\thispagestyle{empty}
		
		\noindent 
		
				\vspace{-25mm}
\singlespacing{\textbf{Abstract:}


By analysing the joint educational distribution of couples in multiple generations, we can learn about the changes in educational homophily and trends in inequality between groups with different income-generating abilities. These data are available for many more countries and decades than individual or household level income data. Therefore, they allow us to document inequality trends in societies and periods that have not previously been analysed with microdata. To study inequality dynamics using couples' data, one needs to apply methods that control for changes across generations in the structural availability of potential partners with various traits.

It is well documented that empirical findings on homophily along with inequality trends in general -- and especially those for America over recent decades -- are very sensitive to the choice of the method applied. Therefore, the method-selection has to be performed with particular care. In this study, we use the Pew Research Center's survey data from years 2010 and 2017 to select the suitable method. The surveys inform us about Americans' self-declared preferences regarding spousal education. The advantage of the analysis performed here over an analysis of survey data from a single survey wave is that it can disentangle the generational and age effects, i.e., it can measure the generation-specific preferences -- that are in the focus of our research -- independently of changes in the survey participants' preferences over their life course.

The results of the analysis confirm the finding of an earlier analysis based on data from a single survey year: namely, that the generation-specific preferences of Americans born after World War II follow a U-shaped pattern. The robustness of the pattern to controlling for the age-effects has the significance that it provides an even stronger basis both for challenging the applicability of a method commonly used until the late 2010s and for supporting a recently proposed alternative method.

}

	\end{titlepage}

\newpage
\setcounter{page}{1}
\doublespacing

\begin{center}
		\Large{\textbf{\large What do surveys say about the trend in inequality and the applicability of two table-transformation methods?}}

\end{center}
{\textbf{Short abstract:}
	
	We use survey data from 2010 and 2017 about Americans' self-reported preferences over spousal education. Unlike a previous analysis using survey data from a single wave, our pseudo-panel analysis allows us to control for changes in preferences over the course of the survey respondents' lives. We test the sign and magnitude of the age-effect-free change in educational homophily from the generation of the early Boomers to the late Boomers, as well as from the early GenerationX to the late GenerationX. We use the test for method selection: to decide whether the conventional iterative proportional fitting algorithm, or its recently proposed alternative is more suitable for analyzing revealed marital preferences. Our test supports the application of the new method and a U-shaped trend in educational homophily in the US. Since the trend in educational homogamy is informative about the dynamics of the overall inequality between different educational groups, the new method, validated with our pseudo-panel analysis of survey data, can be applied to explore trends in inequality pertaining to countries, and periods for which a definitive narrative is lacking so far.

\begin{flushleft}
	\small{\textit{JEL:}  C02, C33, J12.}\\
	\small{\textit{Keywords:}
		Assortative Mating; Iterative Proportional Fitting Algorithm;  NM-method; Pseudo-Panel Analysis; U-shaped Trend in Inequality.}
\end{flushleft}

\newpage

\section{Introduction}

It is a commonly used approach among social scientists to study changes in society by comparing contingency tables of couples from different generations (see \citealp{LichterQian2019}, \citealp{SchwartzMare2005}).   
The joint distributions -- captured by the contingency tables --  reflect revealed marital/
mating preferences at the aggregate level and thereby are informative about which
groups are considered to be in fit socially in each of the generations studied.

To study the trends in the segmentation of the society, one does not only need linked data on couples,  
but also a method suitable for the purpose. As it is shown in the literature of assortative mating and also discussed in this paper,  
the trend in revealed educational homophily, or in the degree of sorting along the educational trait is sensitive to the choice of the method used (see  \citealp{Chiappori_etal2020}, \citealp{Rosenfeld2008}).     
Therefore, it is insightful to compare the trends obtained by various methods on the one hand and the trend in self-reported marital preferences.  
\textit{This paper performs such comparisons through formal tests with the aim of selecting the suitable method. } 

Church marriage registers are dating back several centuries. 
For instance, the data on marriages in England collected and analyzed by \cite{ClarkCummins2022} allow for the study 
of changes in the segmentation of the British society since the industrial revolution. 
However, in this paper we only use data from the relatively recent past.  
Our data are on \textit{American couples} in four generations born after World War II: the \textit{early and late Boomers}, and the \textit{early and late GenXers}. 
More precisely, our data cover the joint educational distributions of couples in these four generations.  

The motivation for focusing on couple formation among Boomers and GenX-ers is three-fold. 
First, unlike generations born much earlier, members of the four generations studied were surveyed about their self-reported views on the importance of spousal education.
Second, we have good quality census data from the late twentieth and early twenty-first centuries not only on official marriages but also on cohabitations.  
Third, the disagreement among different methods about the evolution of revealed preferences is more pronounced when there are such substantial intergenerational changes in the educational distributions of marriageable men and women as we have seen in the case of the Boomers and GenX-ers.

As to the revealed preferences, the challenge of identification is due to the fact that the observed matching outcome depends not only on the \textit{marital homophily}  (or, aggregate marital preferences, or, marital social norms, or, social barriers to marry out of ones' group, or, social gap between different groups, or, segmentation of the marriage market, or, degree of sorting),\footnote{We use these terms interchangeably, because it is hardy possible to distinguish them empirically.} but also on the \textit{structural availability} 
of potential partners 
with different traits (see e.g. \citealp{Kalmijn1998}).  
Therefore, researchers aiming at documenting  
changes in the social divides with marriage data have to net the effect of changing
aggregate preferences from that of the changing structural availability.\footnote{In addition, 
	identifying the changes in preferences is even more challenging once the \textit{possibility of remaining single} and the \textit{possibility of sorting along multiple traits} are also taken into account.   
	About these points, see \cite{NaszodiMendonca2019} and \cite{NaszodiMendonca2023_RACEDU}, respectively.}

For instance, if the assorted trait studied is the education level then  
changes in the social gap among different education groups can be identified from the changes in the share of educationally homogamous couples 
by controlling for the direct and indirect effects of changing education levels of marriageable men and women.

Constructing counterfactuals is the key step of such decompositions since the following two questions cannot be answered without them. First, what would  
be the share of educationally homogamous couples in a certain generation provided women and men in this generation had the same 
education levels as their peers used to have in an earlier born generation. Second, what would  
be the share of educationally homogamous couples in a certain generation provided this generation had the same 
aggregate marital preferences for spousal education as an earlier generation used to have in the past. 

Until the late 2010s, the most popular method for constructing counterfactual joint
distributions in the form of counterfactual contingency tables    
was the table-transformation method called the  \textit{iterative proportional fitting algorithm} (hereafter \href{https://en.wikipedia.org/wiki/Iterative_proportional_fitting}{IPF algorithm}).

The IPF algorithm is a scaling procedure which 
standardizes the marginal distributions of a contingency table   
to a fixed value (where the marginal distributions represent the structural availability), while  
retaining a specific association between the row and the column variables. 
Its preserved association is captured by the odds-ratio if the assorted trait is dichotomous.

``It is impossible, or not easy, to alter by argument what has long been absorbed by habit''.\footnote{See: Nicomachean Ethics X.1179b.} 
Despite of Aristotle's admonition, we dare to argue in this paper that the habit of many researchers to construct counterfactuals with the \textit{IPF has a better alternative}.  
In particular,  we show that the table-transformation method developed by \cite{NaszodiMendonca2021} (henceforth \href{https://en.wikipedia.org/wiki/NM-method}{NM-method}) is more suitable for the purpose than the IPF.\footnote{While the IPF is a built-in algorithm in SPSS, the NM is not yet.  
	However, there is no technical obstacle for the NM-method to gain popularity among non-SPSS users    
	since it is implemented in R, Excel,  Visual Basic, Matlab, and Stata   
	(see: \url{http://dx.doi.org/10.17632/x2ry7bcm95.2} and \url{https://data.mendeley.com/datasets/95k6mmrxvg}).}  
The transformed table obtained with the NM has the same preset marginal distributions as the transformed table obtained with the IPF. 
However, the NM-transformation is invariant to the ordinal indicator proposed by \cite{LiuLu2006} (henceforth, LL-indicator), rather than being invariant to the cardinal odds-ratio.

The focus of this paper is on the empirical performance of the IPF-based and NM-based counterfactual decompositions at quantifying the dynamics of a certain dimension of inequality.  
This dimension of inequality is considered to increase (/decrease) between different education groups if  
the aggregate preferences for educational homogamy or well-educated partners are found to be stronger (/weaker) in a later generation relative to an earlier generation.\footnote{As it is pointed out by   \cite{Becker1973}, 
	\cite{Hitsch2010} and \cite{Kalmijn1998}, the preferences for educational homogamy and the preferences for well-educated partners are empirically equivalent provided only the matching outcome is observed. In other words, one cannot distinguish between horizontal and vertical preferences using couples' joint distributions only. This point motivates us to use survey data, informative about the latter type of preferences, for selecting the method suitable for analyzing the former type of preferences.}

As to the \textit{empirical results}, \cite{NaszodiMendonca2021} get {qualitatively different findings by applying the IPF and the NM} for counterfactual decompositions using US census data. In particular, by studying the revealed aggregate preferences of young American adults with the {IPF}, the American late Boomers are found to be \textit{more}   ``picky'' about spousal education than the early Boomers. By contrast, the {NM} supports the opposite result.

Regarding the \textit{quantitative} results, the IPF suggests that the share of educationally homogamous young couples would have \textit{increased by one percentage point} if only the marital preferences had changed from the generation of the early Boomers to the generation of the late Boomers.
By contrast, the NM quantifies the same ceteris paribus effect to 
have the opposite sign as being  \textit{minus three  percentage points} (see the dark bars in Figure \ref{fig:SHC_19802010} for the period  1980--1990).

\begin{figure}
	\begin{center}
		\caption{Decomposition of changes in the share of educationally homogamous couples between 1980 and 2010 in the US -- male partners are 30--35 years old}
		\centering
		\includegraphics[width=1\linewidth]{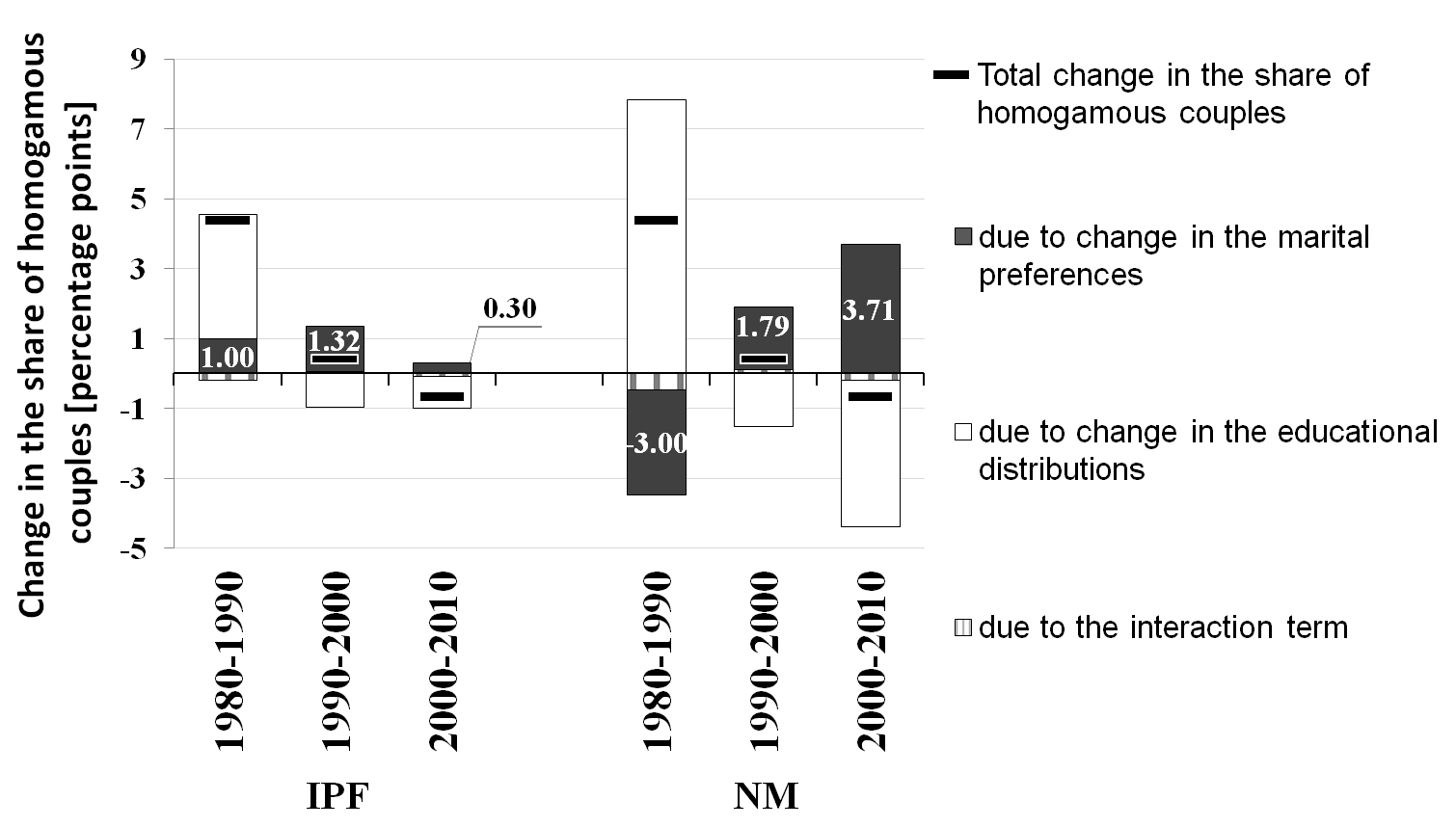}
		\label{fig:SHC_19802010}\\
	\end{center}
	\textit{Notes}: Partial replication of Figure 1b in \cite{NaszodiMendonca2021}. They use \textit{census data} from IPUMS on both the education level of married couples and cohabiting couples in 1980, 1990, 2000 and 2010. 
	The male partners aged 30--35 years in 1980, 1990, 2000 and 2010 belong to the generations of the early Boomers, the late Boomers,  
	the early GenerationX and the late GenerationX, respectively. The educational attainment can take three different values: ``Low'': no high school degree, ``Medium'': high school degree without tertiary level diploma, and ``High'': tertiary level diploma. Accordingly, the educationally homogamous couples are those, where both the female partner and the male partner are either ``Low'' educated, or both are ``Medium'' educated, or both are ``High'' educated.      
	Changes in the prevalence of homogamy across consecutive generations  are  decomposed by using the so called {additive decomposition scheme with interaction effects} proposed by \cite{Biewen2014}, while the counterfactual tables are constructed either by the IPF, or the NM.  
	
\end{figure}

The two methods also disagree on whether  educational homophily  was  \textit{much more} intense   
in the late GenerationX  than in the early GenerationX, or just \textit{slightly more} intense.\footnote{While the decomposition makes this disagreement on the magnitudes  explicit over the marital/ mating preferences in the GenerationX, the disagreement is obscured by the direct comparison of the odds-ratio and the LL-measure due to the ordinal nature of the latter.}     
Quantitatively, the NM attributes \textit{almost four percentage points 
	increase} in the share of educationally homogamous couples to the changing homophily.       
Whereas the IPF quantifies the same effect to be \textit{close to zero} (0.3 percentage point). 
These results  can be read from the height of the dark bars in Figure \ref{fig:SHC_19802010} corresponding to the period  2000--2010.

By exploiting the above disagreements between the methods, \cite{NaszodiMendonca2021} perform an \textit{empirical method-selection}.  
For the selection, they present some descriptive statistics of a survey conducted by the Pew Research Center in 2010.     
They find that the difference between the survey responses of the  early Boomers and the late Boomers, 
as well as the difference between the responses of the members of the early GenerationX and the late GenerationX,  
support the application of their method.  
In particular, \textit{much less} late Boomers than early Boomers agreed with a statement about the importance of being well-educated by the members of the opposite sex in order to become a good spouse. 
And \textit{much more} respondents agreed with the same statement from the late GenerationX than from the early  GenerationX.

The \textit{contributions} of this paper relative to the study by \cite{NaszodiMendonca2021} are three-fold.   
First, in this paper, we present not only some survey evidence supporting the NM in the context of assortative mating, but also some \textit{general theoretical considerations} relevant for other applications as well. 
Second, unlike the study by \cite{NaszodiMendonca2021}, this paper performs some \textit{formal hypothesis tests}. 

Last, but not least, we complement the preliminary survey analysis by \cite{NaszodiMendonca2021} in an important respect.   
Unlike Naszodi and Mendonca, we use \textit{survey data from two waves}.  
Using a richer set of data is {motivated} by the following point.   
Although the survey data from a single wave are informative about the variations of preferences of individuals belonging to different generations, these individuals are also different in terms of their age. 
If the self-reported marital preferences can vary substantially over the course of individuals' lives   
then the generation-effects one can identify from a single wave of a survey are not net of a potentially major confounding factor, i.e., the age-effect.

Our pseudo-panel analysis of survey data from two waves allows us to identify the \textit{net generation-effects capturing what the differences between the generation-specific responses would be like if the respondents from different generations were interviewed at the same age}. We study the responses of men and women survey participants separately and test the following four \textit{hypotheses}:\\ 
(i) the share of ``picky'' \textit{men}  would have been lower in the population of the \textit{late Boomers} than in the population of the \textit{early Boomers} if every men  in these two generations had been interviewed at the same age;\\   
(ii) the share of ``picky'' \textit{women}  would have been  lower in the population of the \textit{late Boomers} than in the population of the \textit{early Boomers} if every women  in these two generations had been interviewed at the same age;\\   
(iii)  the share of ``picky'' \textit{men} would have been much higher among the \textit{late GenerationX} than among the \textit{early GenerationX} if every men in these two generations had been interviewed at the same age;\\
(iv)  the share of ``picky'' \textit{women} would have been much higher among the \textit{late GenerationX} than among the \textit{early GenerationX} if every women in these two generations had been interviewed at the same age. 

Depending on the results of these tests, the survey data validate either  the application of the NM, or the IPF (or neither)  for constructing counterfactuals.

The structure of this paper is the following. 
In Section \ref{sec:theory}, we present some theoretical considerations 
concerning the applicability of the  IPF and the  NM. 
Section \ref{sec:datamethod} introduces the data and the method used for the hypothesis tests before it presents the results of the tests.    
In Section \ref{sec:disc}, we discuss the significance of the empirical findings.   
Finally, Section \ref{sec:concl} concludes the paper.

\section{Some theoretical considerations concerning the applicability of the IPF and the NM}\label{sec:theory}

This section provides a formal definition of the table-transformation methods compared, i.e., the IPF and the NM.  
Also, it presents some {theoretical considerations concerning their applicability}.  

In particular, \textit{we argue that the IPF and the NM provide solutions for two different sets of problems:  
while the IPF is suitable for completing a population table by using a sample from the population, the 
NM is fit for constructing a counterfactual population table from two population tables.}  

In addition, we illustrate with two numerical examples that the transformed tables obtained with
the IPF can be sensitive to the choice of the number of trait categories. 
Therefore, the outcomes of certain counterfactual decompositions are also sensitive to the same choice provided the counterfactual tables are constructed by the IPF rather than the NM.

Let us start with the definitions that we will illustrate with the examples of two pairs of contingency tables.   
The pair of contingency tables, denoted by $P$ and $Q$, are of the same size and they representing joint distributions of two categorical traits.  
E.g., $P$ and $Q$ can represent the joint distribution of  husbands' and wives' education levels in an earlier born generation and a later born generation, respectively.      
In an alternative example, 
$P$  represents the joint distribution of objects in a box by material and shape, while  
$Q$  represents the joint distribution of objects in a random sample taken from the box.   
The two examples are illustrated by Tables \ref{tab:CT}  and \ref{tab:CT2} showing 2-by-2 $P$ and $Q$ tables together with their row sums and column sums. 

\begin{table}[ht]
		\caption{First example for the contingency tables $P$ and $Q$ representing joint educational distributions of couples in two generations}

\begin{minipage}[p]{0.48\columnwidth}
	\centering
	\begin{tabular}{llcc c}  
		\hline \hline
	\multirow{2}{*}{$P$} 		& 	 & \multicolumn{3}{c}{Wives in earlier generation}   	\\ \cline{3-5}
	&	& $L$  & $H$ & Sum \\ \cline{1-5} 
	\multirow{2}{*}{\rotatebox[origin=c]{90}{{\small Husb.}}}	& $L$  & $N^{\text{early}}_{L,L}$  & $N^{\text{early}}_{L,H}$  & $N^{\text{early}}_{L,\cdot}$ \\ 
	& $H$ & $N^{\text{early}}_{H,L}$ & $N^{\text{early}}_{H,H}$   & $N^{\text{early}}_{H,\cdot}$ \\   \cline{2-5}
	& Sum & $N^{\text{early}}_{\cdot,L}$ & $N^{\text{early}}_{\cdot,H}$  & $N^{\text{early}}$ \\ 	 \cline{1-5} \hline \hline
\end{tabular}

\end{minipage}
\hfill
\begin{minipage}[p]{0.48\columnwidth}
	\centering
	\begin{tabular}{llcc c}  
			\hline \hline
		
	\multirow{2}{*}{$Q$} 		& 	 & \multicolumn{3}{c}{Wives in later generation}   	\\ \cline{3-5}
	&	& $L$  & $H$ & Sum \\ \cline{1-5} 
	\multirow{2}{*}{\rotatebox[origin=c]{90}{{\small Husb.}}}	& $L$  & $N^{\text{late}}_{L,L}$  & $N^{\text{late}}_{L,H}$  & $N^{\text{late}}_{L,\cdot}$ \\ 
	& $H$ & $N^{\text{late}}_{H,L}$ & $N^{\text{late}}_{H,H}$   & $N^{\text{late}}_{H,\cdot}$ \\   \cline{2-5}
	& Sum & $N^{\text{late}}_{\cdot,L}$ & $N^{\text{late}}_{\cdot,H}$  & $N^{\text{late}}$ \\ 	 \cline{1-5}
	\hline \hline
	
\end{tabular}

\end{minipage}
\begin{tablenotes}
	\small
	\item \textit{Note}: $L$ and $H$ denote low and high levels of education, respectively.
	
\end{tablenotes}

\label{tab:CT}
\end{table}


	\begin{center}
\end{center}

\newpage

\begin{table}[ht]
	\caption{Second example for the contingency tables $P$ and $Q$ representing joint distributions of objects by material and shape in a box and a random sample of objects taken from the box}
\begin{minipage}[p]{0.48\columnwidth}
	\centering

\begin{tabular}{llcc c}  
	\hline \hline
	
	\multirow{2}{*}{\textit{P}} 	 &	 & \multicolumn{3}{c}{Shape of objects in a box}   \\ \cline{3-5}
	&	& Tetrahedron  & Cube & Sum \\ \cline{1-5} 
	\multirow{3}{*}{\rotatebox[origin=c]{90}{{\small Material}}} &	Argent  & $N_{{\text{A}},{\text{T}}}$ & $N_{{\text{A}},{\text{C}}}$  & $N_{{\text{A}},\cdot}$ \\ \\ 
	&Gold & $N_{{\text{G}},{\text{T}}}$ & $N_{{\text{G}},{\text{C}}}$  & $N_{{\text{G}},\cdot}$ \\  \cline{2-5}
	&Sum & $N_{\cdot,{\text{T}}}$ & $N_{\cdot,{\text{C}}}$  & $N_{\cdot,\cdot}$ \\ 	 \cline{1-5} 	\hline \hline
	
\end{tabular}
\end{minipage}
\hfill
\begin{minipage}[p]{0.48\columnwidth}

\centering
\begin{tabular}{llcc c}  
		\hline \hline
	
	\multirow{2}{*}{\textit{Q}} 	 &	 & \multicolumn{3}{c}{Shape of objects in a sample}   \\ \cline{3-5}
	&	& Tetrahedron  & Cube & Sum \\ \cline{1-5} 
	\multirow{3}{*}{\rotatebox[origin=c]{90}{{\small Material}}} &	Argent  & $n_{{\text{A}},{\text{T}}}$ & $n_{{\text{A}},{\text{C}}}$  & $n_{{\text{A}},\cdot}$ \\ \\
	&Gold & $n_{{\text{G}},{\text{T}}}$ & $n_{{\text{G}},{\text{C}}}$  & $n_{{\text{G}},\cdot}$ \\  \cline{2-5}
	&Sum & $n_{\cdot,{\text{T}}}$ & $n_{\cdot,{\text{C}}}$  & $n_{\cdot,\cdot}$ \\ 	 \cline{1-5} \hline \hline
	
\end{tabular}

\end{minipage}
\begin{tablenotes}
	\small
	\item \textit{Note}: $A$ and $G$ represent argent and gold, while $T$ and $C$ represent tetrahedrons and cubes, respectively.
	
\end{tablenotes}


\label{tab:CT2}
\end{table}

The IPF algorithm applied to tables $P$ and $Q$ is defined by the following two steps to be iterated until convergence.  
First, it factors the rows of the seed table $Q$ in order to match the row totals of $P$.  
In our first example, this step involves multiplying $Q$ with the transpose of  $\begin{bmatrix}
	N^{\text{early}}_{L,\cdot}/N^{\text{late}}_{L,\cdot} & N^{\text{early}}_{H,\cdot}/N^{\text{late}}_{H,\cdot}    
	\end{bmatrix}$. 
	In the alternative example, this step involves multiplying $Q$ with the transpose of  $\begin{bmatrix}
	N_{{\text{A}},\cdot}/n_{{\text{A}},\cdot} & N_{{\text{G}},\cdot}/n_{{\text{G}},\cdot}    
	\end{bmatrix}$.

The table obtained after the first step (to be denoted by $Q^{'}$) may not have its column totals equal to the column totals of $P$. 
In this case, it is necessary to perform a second step.         
As the second step, the IPF factors the columns of $Q^{'}$ to match the corresponding column totals of $P$. 
In our first example, the second step involves multiplying  $\begin{bmatrix}
N^{\text{early}}_{\cdot,L}/N^{\text{late}}_{\cdot,L} & N^{\text{early}}_{\cdot,H}/N^{\text{late}}_{\cdot,H}    
\end{bmatrix}$ with $Q$. 
In the alternative example, the second step involves multiplying  $\begin{bmatrix}
N_{\cdot,{\text{T}}}/n_{\cdot,{\text{T}}} & N_{\cdot,{\text{C}}}/n_{\cdot,{\text{C}}}    
\end{bmatrix}$ with $Q$.

The table obtained  after  this  step (to be denoted by $Q^{''}$) may not have its row totals equal to the row  totals of $P$. 
In this case, repeating the first step is necessary with $Q=Q^{''}$. 
Alternatively, we stop the iteration. 

The table constructed by the IPF is the last value of $Q^{''}$ before we stop the iteration.   
We denote it by $Q^{\text{IPF}}$.  
Let us visit some of its properties.  
Table $Q^{\text{IPF}}$  has row totals that match the row totals of $P$, also  its column totals match the  column totals of $P$ by construction.  
More importantly, if $Q$ and $P$ are 2-by-2 tables, then the odds-ratio of $Q^{\text{IPF}}$ is the same as that of table $Q$, because the odds-ratio's 
numerator and denominator are multiplied by the same scalar in each step of the iterative procedure. 
For instance, the odds-ratio of $Q$ is $N^{\text{late}}_{L,L} N^{\text{late}}_{H,H}/(N^{\text{late}}_{L,H}N^{\text{late}}_{H,L})$ in the first example. 
In the same example, the odds ratio of 
$Q^{\text{IPF}}$ is $N^{\text{late}}_{L,L} N^{\text{late}}_{H,H}/(N^{\text{late}}_{L,H}N^{\text{late}}_{H,L}) \times  (N^{\text{early}}_{L,\cdot}/N^{\text{late}}_{L,\cdot})^{2k}/(N^{\text{early}}_{L,\cdot}/N^{\text{late}}_{L,\cdot})^{2k}  \times (N^{\text{early}}_{H,\cdot}/N^{\text{late}}_{H,\cdot})^{2k}/ (N^{\text{early}}_{H,\cdot}/N^{\text{late}}_{H,\cdot})^{2k}=N^{\text{late}}_{L,L} N^{\text{late}}_{H,H}/(N^{\text{late}}_{L,H}N^{\text{late}}_{H,L})$, where k is the number of iterations.\footnote{This property of the IPF is not new. It was already pointed out by \cite{Fienberg1970}.}

The IPF was invented by \cite{StephanDeming1940}. 
It is important to note that they      
developed the IPF with a purpose different from constructing counterfactuals.  
Their purpose was to 
estimate a contingency table of a \textit{population}  from its known marginal distributions   
and a known 
contingency table of a random \textit{sample} from the same population.  
In this exercise of \textit{``completing a population table by using a sample''},      
both the marginal distributions and  
the sample characterize the same population (e.g. the population of objects in a box); moreover,  
the marginal distributions and the sample are observed at the same time.  

Our first theoretical point related to the applicability of the IPF is this.   
The set of problems of {``completing a population table''} is different from the set of problems of {``constructing a counterfactual table''}.
In contrast to the problems of {``completing a population table''},  
in the  exercises of \textit{``constructing a counterfactual table''}, 
the marginal distributions and the seed table characterize two different populations (e.g., the populations of two generations).  
Accordingly, the table constructed represents a \textit{counterfactual prediction}.  
For instance, in the context of assortative mating, it can represent a prediction on what the marriage patterns would be like in a generation    
if only the preferences  had changed relative to an earlier generation, but not the structural availability.\footnote{Alternatively, if the marginal distributions and the seed table characterize the populations of two different societies observed simultaneously, e.g. Americans and UK citizens observed this year, then the counterfactual table can represent a prediction on what the marriage patterns would be like in the US, if the educational distributions of marriageable American men and women were preserved, while they would be sorted into couples just like the British.} 

Our second point is about the equivalence of the IPF and the maximum likelihood estimator and their applicability.    
In the  {``completing a population table''} exercise, the maximum likelihood estimator is a natural candidate for estimating the missing values in the \textit{population} table by using  data in a random \textit{sample}.\footnote{For instance, the maximum likelihood is applicable to estimate the missing values of $N_{{\text{A}},{\text{T}}}$, $N_{{\text{A}},{\text{C}}}$, $N_{{\text{G}},{\text{T}}}$, $N_{{\text{G}},{\text{C}}}$ from their sample counterparts $n_{{\text{A}},{\text{T}}}$, $n_{{\text{A}},{\text{C}}}$, $n_{{\text{G}},{\text{T}}}$, $n_{{\text{G}},{\text{C}}}$ and $N_{\cdot,{\text{T}}}$, $N_{\cdot,{\text{C}}}$, $N_{{\text{A}},\cdot}$, $N_{{\text{G}},\cdot}$ in our second example.} 
However, nothing justifies the application of the  maximum likelihood for estimating a population table from two {population} tables.\footnote{Conversely, nothing justifies the application of the NM for completing a population table by using a sample table. For this reason -- and contrary to the comment of an anonymous reviewer -- the NM cannot be validated by the goodness of fit of its out-of-sample prediction.}     
So, nothing justifies the application of the maximum likelihood for constructing a counterfactual prediction from the observations of two populations.           
Actually, the IPF constructs the same table as the maximum likelihood estimator (see \citealp{Meyer1980}).   
Therefore, there is no guarantee that the IPF would fit for estimating counterfactual tables either, even if it is perfectly suitable for the purpose it was originally developed for.

Our third theoretical point is this. 
Using the IPF to construct  a counterfactual prediction in the context of assortative mating implicitly assumes 
that the aggregate marital preferences, or the degree of sorting in the two generations are characterized 
by the kind of association the IPF preserves, while it transforms the seed table.     
In other contexts, the applicability of the IPF relies on the same kind of assumption as the one in the context of assortative mating: 
the association preserved is assumed to be exactly the one we want to control for.      
Stephan and Deming recognized the crucial role of this assumption.\footnote{In the context of assortative mating,  the related assumptions are difficult to test since marital preferences are rarely observed directly.}
They even warned that  their  algorithm is ``not by itself useful for prediction''  (see \citealp{StephanDeming1940} p.444).

Finally, we present two numerical examples for the application of the IPF with ordered assorted traits.   
These examples illustrate the point that \textit{the transformed tables obtained with the IPF can be sensitive to the choice of the trait categories due to an unfortunate analytical  property of the IPF.}  
The analytical property is this:  the IPF does not commute with the operation of merging neighboring categories of the assorted traits. 
The related sensitivity can be especially problematic if the ordered trait variables (e.g. the husbands' and the wives' education levels)  are available at a relatively high granularity. 
It allows the researcher to manipulate the result of the counterfactual decompositions performed with the IPF, even unconsciously. 

In addition, such sensitivity -- if it is not recognized to be simply due to a poor analytical property, i.e., the lack of commutativity of some measures commonly applied -- can support the misconception that the empirical assortative mating literature is not conclusive about a number of research questions. 
However, if we disregard all the findings obtained using measures and methods with poor analytical properties, many of our empirical findings will become much less puzzling.

In our \textit{first numerical example}, the assorted traits are dichotomous (e.g., both the husbands' and the wives' education level can take the values $L$ and $H$). 
Accordingly, $P$ and $Q$ are 2-by-2 contingency tables:  
$P^{\text{num1}}=$ 
\scalebox{0.7}{$\begin{bmatrix}
	500   & 700 \\
	100   & 700
	\end{bmatrix}$} 
and 
$Q^{\text{num1}}= {\scriptsize \begin{bmatrix}
	500   & 500  \\
	100   & 900
	\end{bmatrix} }$. 

As a first step of the IPF algorithm, we factor the rows of the seed table $Q^{\text{num1}}$ in order to match the row totals of $P^{\text{num1}}$.  
The table obtained after  the first step is $Q^{' \text{num1}}= {\scriptsize \begin{bmatrix}
	600   & 600 \\
	\;80   & 720
	\end{bmatrix} }$. 
As a second step, we factor the columns of  $Q^{'\text{num1}}$ in order to match the column totals of $P^{\text{num1}}$.  
We get  $Q^{''\text{num1}}= {\scriptsize \begin{bmatrix}
	529.41   & 636.36 \\
	\;70.59   & 763.64
	\end{bmatrix} }$. After 4  iterations, the (rounded) IPF-transformed table is  
$Q^{\text{IPF},\text{num1}}= {\scriptsize \begin{bmatrix}
	534	& 665 \\
	\;66	& 735
	\end{bmatrix} }$.  

In our \textit{second numerical example}, one of the assorted traits is dichotomous, while the other one is trinomial  
(e.g. the husbands' trait can take the values low and high, while the wives' ordered trait can  take the values low, medium and high). 
Tables $P$ and $Q$ are given by the following 2-by-3 tables:      
$P^{\text{num2}}= {\scriptsize \begin{bmatrix}
	500   & 400 &  300 \\
	100   & 400 & 300
	\end{bmatrix} }$ 
and 
$Q^{\text{num2}}= {\scriptsize \begin{bmatrix}
	500   & 300 &  200 \\
	100   & 300 & 600
	\end{bmatrix} }$. 
After 4 iterations, the (rounded) IPF-transformed table is  
$Q^{\text{IPF},\text{num2}}= {\scriptsize \begin{bmatrix}
	528   & 475 &  197 \\
	\;72   & 325 & 403
	\end{bmatrix} }$.
By merging its last two columns, we get 
$Q^{\text{IPF},\text{num2},\text{merged}}= {\scriptsize \begin{bmatrix}
	528	& 672 \\
	\;72	& 728
	\end{bmatrix} }$.  
Apparently, this table is different from  $Q^{\text{IPF},\text{num1}}$ despite the fact that 
$P^{\text{num1}}$ and $Q^{\text{num1}}$ are equal to $P^{\text{num2}}$ and $Q^{\text{num2}}$ with merged last two columns, respectively.  
The difference between $Q^{\text{IPF},\text{num1}}$  and $Q^{\text{IPF},\text{num2},\text{merged}}$  does not vanish if we continue iterating after the fourth step.

If the tables in the above examples represent joint educational distributions of
couples in two generations, then tables  
$Q^{\text{IPF},\text{num1}}$ and 
$Q^{\text{IPF},\text{num2},\text{merged}}$ suppose to represent 
the joint educational distributions of couples under a counterfactual. 
In particular, the counterfactual is that 
aggregate marital preferences are the same as in the generation represented by table $Q$, while the 
educational distributions of marriageable men and marriageable women are the same as in 
the generation represented by table $P$.    

The difference between 
$Q^{\text{IPF},\text{num1}}$ and 
$Q^{\text{IPF},\text{num2},\text{merged}}$
makes a difference in the outcome of a counterfactual 
decomposition. 
Out of the $\frac{500+900}{2,000}-\frac{500+700}{2,000}=10$ 
percentage points increase in the shares of educationally homogamous couples
from the generation represented by table $P^{\text{num1}}$  to the generation
represented by table $Q^{\text{num1}}$, we attribute either
$\frac{534+735}{2,000}-\frac{500+700}{2,000}=3.45$ percentage points
difference, or
$\frac{528+728}{2,000}-\frac{500+700}{2,000}=2.8$ percentage points
difference 
to the changing aggregate preferences from one generation to the other. 
So, the result of our decomposition is sensitive to whether we choose not to distinguish between the last two neighboring educational categories before constructing the counterfactual table with the IPF, or after it.  

We make the remark that, in fact, the number of categories used varies across the empirical papers in the educational assortative mating literature.   
For instance, \cite{ChooSiow2006}, as well as \cite{NaszodiMendonca2021}, distinguish between three educational categories (``no high school degree'', ``high school graduates'', and ``college graduates''). Whereas \cite{Eika2019} divide the middle category to ``high school degree with no college'' and ``high school degree with some college''.  \cite{GihlebLang2020} use five, six and twelve categories. They propose to rely on the wage structure literature (see \citealp{acemoglu2011}) when selecting the educational categories.\footnote{We warn that the number of categories should be chosen carefully even if the method used for constructing the counterfactuals commutes with the operation of merging neighboring categories. 
  As an example for not careful selection of categories, imagine that we analyze matching along age (or any other trait described by a continuous variable) and a couple is considered as being homogamous if 
	their age difference is below a certain threshold. Provided the number of age categories is chosen to be extremely high with a threshold as low as one minute, the share of homogamous couples is close to zero. In addition, contrary to common sense, the share of homogamous couples is practically unchanged across any pair of consecutive generations under such an  extremely granular set of age categories.}

None of the above points rule out that the IPF can perform well in some empirical applications at constructing certain counterfactuals.      
However, these points show that the choice of the IPF is not sufficiently justified theoretically.      
Our points call for a counterfactual table constructing method that is \textit{theoretically appealing}, while it can also be \textit{validated empirically}.   

Recently, \cite{NaszodiMendonca2021} proposed a table-transformation method, the NM-method,  as an alternative to the IPF. 
The NM is defined as the  method that transforms table $Q$ into another table with preset row sums and column sums determined by table $P$, while retaining a specific association between the row and the column variables.  
Its preserved association is captured by the scalar-valued LL-indicator (rather than the odds-ratio) if $Q$ and $P$ are 2-by-2 tables.  
For larger tables, the retained association is captured by the matrix-valued generalized LL-measure (see \citealp{NaszodiMendonca2021} for the definition of the generalized LL-measure).

\cite{NaszodiMendonca2021} visit an important property of the NM. 
Namely, that it commutes with the operation of merging neighboring categories of the ordered assorted traits by construction. 
The significance of commutativity is that social scientists applying the NM cannot directly influence the outcome of the counterfactual decompositions by choosing the granularity of the categorical variables whose joint distribution is studied. 
For instance, in our numerical examples, the NM-transformed table is     
$Q^{\text{NM,}{\text{num1}}}=Q^{\text{NM,}{\text{num2}}, \text{merged}}=  {\scriptsize \begin{bmatrix}
	520   & 680 \\
	\;80   & 720
	\end{bmatrix} }$ 
irrespective of first merging the last two columns of $P^{\text{num2}}$ and $Q^{\text{num2}}$  before applying the NM, or the other way around (see the online Appendix A).

 As to the empirical validation of the NM,  \cite{NaszodiMendonca2021} present some survey evidence in favor of its  application in the context of analyzing changes in assortative mating by counterfactual decompositions. In the next section, we refine their validation exercise.

\section{Testing hypotheses with survey data about self-reported preferences}\label{sec:datamethod}

For the analysis, we use not only the Pew Research Center's survey called \textit{Changing American Family Survey} from 2010 (that was used by \citealp{NaszodiMendonca2021}), but also  its    																						
\textit{American Trends Panel Wave 28 Survey} conducted in 2017. 

In 2017, almost the same pair of questions (coded QUALHUSB and QUALWIFE) was asked as in 2010 (coded Q.23F1 and Q.24F2).  
The question 
asked from female survey participants in 2017 was: ``How important, if at all, do you feel it is for a good \textit{husband} or partner to be well educated?''. 
The question 
asked from male survey participants in 2017 was: ``How important, if at all, do you feel it is for a good \textit{wife} or partner to be well educated?''. 
The potential responses offered for the participants were: (i) very important; (ii) somewhat important; (iii) not too important; (iv) not at all important; (v)	don't know/ refuse to answer.  

In 2010, the question Q.23F1 (/Q.23F2)
asked from female (/male) participants was the following. ``People have different ideas about what makes a man (/woman) a good husband (/wife) or partner. For each of the qualities that I read, please tell me if you feel it is very important for a good husband (/wife) or partner to have, somewhat important, not too important, or not at all important. First, he (/she) is well educated. Is this very important for a good husband (/wife) or partner to have, somewhat important, not too important, or not at all important?''  
Apparently,  the formulation of the questions in 2010 and 2017 were just slightly different, while the potential responses were exactly the same.   

To make our results comparable with the results in \cite{NaszodiMendonca2021}, we use survey data from 2010 on the same generations as they do. 
In particular, the early Boomers are represented in our analysis by those who were born between 1946 and 1950;  
the  late Boomers are represented by those who were born between 1956 and 1960;  
the   early GenerationX is represented by those who were born between  1966 and 1970;  
the  late  GenerationX is represented by those who were born between 1976 and 1980. 
With this choice, we sidestep the problem of determining the exact year of birth of the last members of the generations studied and that of the first members of the next generations.

Although many of the survey participants were interviewed in both of the survey waves, our survey data is not in a panel structure.  
In fact, the Pew Research Center shared publicly the survey data in 2017 by using broader age categories than in 2010. 
In particular, while the year of birth is reported in the 2010 survey data, we know only if the survey respondents in 2017 belonged to the GenerationX (i.e., belonged to the age group populated by the  37--52 years old individuals) or belonged to the Boomer generation (i.e., belonged to the  53--71 years old cohort).\footnote{The \textit{distributions of survey participants} by gender, generation  and also by the dichotomous type of respondent--non-respondent are presented in 
	 Appendix B.}    
This variation of the age categories across the two survey waves makes it impossible to  link the survey participants and to construct a panel.      

Still, we can apply a \textit{pseudo-panel analysis} by  
collecting and comparing the answers given in 2010 and 2017 of    
those Boomers, who were born between 1946 and 1964. 
Also, we can compare the answers given in 2010 and 2017 by the members of the
GenerationX born between 1965 and 1980.      
Thereby, we can quantify and control for the age-effects: how the responses of the
Boomers and those of the GenerationX have changed over seven years between
2010 and 2017.

Let us see the steps of our pseudo-panel analysis.   
First, we estimate the \textit{population-shares} of those men and women, who viewed spousal education to be very important.  
For the estimation, we apply the approximation proposed by \cite{AgrestiCoull1998} (see Equation \ref{eq:AC}).    
Our estimated population-shares are not only gender-specific, but also generation-specific and period-specific since the shares vary over one generation to another and also over the survey waves.     
Accordingly, we denote the estimated population-shares by $\widehat{PS}_{\text{gender}, \text{generation}, t}$.

Second, we calculate the \textit{generation-effects with the confounding age-effects} by using  data from 2010 exclusively. 
It is calculated as the difference between the estimated population-shares in the two generations to be compared.
We denote it by ${GE}_{\text{gender}, \text{generations}, t}$.  
For the late Boomers and the early Boomers, we estimate it as 
\begin{equation}
\widehat{GE}_{\text{gender}, \text{late and early Boomers}, 2010}= \widehat{PS}_{\text{gender}, \text{late Boomers}, 2010}-\widehat{PS}_{\text{gender}, \text{early Boomers}, 2010} \;.
\end{equation} 
Similarly, for the late GenerationX and the early GenerationX to be compared, we estimate the generation-effect by 
\begin{equation}
\widehat{GE}_{\text{gender}, \text{late and early GenX}, 2010}= \widehat{PS}_{\text{gender}, \text{late GenX}, 2010}-\widehat{PS}_{\text{gender}, \text{early GenX}, 2010} \;.
\end{equation}

Third, we calculate the \textit{age-effects}, i.e., how the responses of the Boomers  and those in the  GenerationX  have changed over seven years between 2010 and 2017.
Similarly to the population-shares, the age-effects are also gender-specific, generation-specific and  period-specific.      
Accordingly, we denote it by ${AE}_{\text{gender}, \text{generation}, t, \Delta t}$ and estimate it as 
\begin{equation}
\widehat{AE}_{\text{gender}, \text{generation}, 2010, 7}= \widehat{PS}_{\text{gender}, \text{generation}, 2017}-\widehat{PS}_{\text{gender}, \text{generation}, 2010} \;.
\end{equation}

Fourth, we calculate the \textit{net generation-effects} by adjusting our biased estimates on the generation-effects obtained in step two.   
Since the average age difference between the early and late Boomers, as well as between the early and  late GenerationX, is 10 years,   
we have to perform the adjustments with the kind of age-effects that capture how the responses of a studied generation change over 10 years.        
We assume that the responses of the Boomers and those in the GenerationX would have changed over 10 years as much as  10/7 times the age-effects identified in step three. 
Accordingly, we estimate the net generation-effect for the Boomers as   
\begin{equation}
\widehat{NGE}_{\text{gender}, \text{Boomers}, 2010, 2020}=\widehat{GE}_{\text{gender}, \text{late and early Boomers}, 2010} +10/7 \widehat{AE}_{\text{gender}, \text{Boomers}, 2010, 7} \;. 
\end{equation} 
Whereas for the GenerationX, the net generation-effect is estimated by  
\begin{equation}
\widehat{NGE}_{\text{gender}, \text{GenX}, 2010, 2020}=\widehat{GE}_{\text{gender}, \text{late and early GenX}, 2010} +10/7 \widehat{AE}_{\text{gender}, \text{GenX}, 2010, 7} \;. 
\end{equation}

Finally, we construct the confidence intervals around the point estimates.  
For the population-share, we rely again on the work by \cite{AgrestiCoull1998}. 
Following them, we assume that the distribution of the number of survey responses 
``very important'' (denoted by $x$) out of  $n$ number of total responses is binomial with the parameter $PS$. 
While $q=x/n$ is the sample-share of ``picky'' respondents,  $PS$ denotes their population-share.  
\cite{AgrestiCoull1998} propose to estimate the latter as  
\begin{equation}\label{eq:AC}
\widehat{PS}=(x+z^2/2)/(n+z^2) \;, 
\end{equation} 
where $z=\Phi^{-1}(1-\alpha/2)$ is the quantile of a standard normal distribution. 
(For example, a 95\% confidence interval requires $\alpha=0.05$, thereby producing $z=1.96$.)  
And they propose to approximate the population-share's symmetric confidence interval by  
\begin{equation}\label{eq:ACconf}
\widehat{PS} \pm z \widehat{\sigma}_{PS} \;, 
\end{equation} 
where  $\widehat{\sigma}_{PS}= \sqrt{\widehat{PS}(1-\widehat{PS})/(n+z^2) }$ is the estimates for the standard error. 
(For the sake of simplicity, we omitted the indices of both  $\widehat{PS}$ and  $\widehat{\sigma}_{PS}$. 
However, in our empirical application, both are gender-specific, generation-specific and period-specific.)

To implement the empirical analysis, we have to set the value of a kind of correlation that we denote by $\rho$.     
This correlation captures to what extent the response given in 2017 by the representative survey participant resembles the response of the same person given in 2010.  
We calibrate $\rho$ equal to zero in our benchmark analysis, although it is not reasonable to assume the responses to be uncorrelated. 
As we will see, by calibrating $\rho$ to the value zero, we get maximally conservative test results.    
While the point estimates of $NGE$ is independent of  $\rho$, its confidence interval is not.  

The symmetric \textit{confidence intervals} of the generation-effects, age-effects and net generation effects are computed similarly to the confidence interval of the population-share: the upper bound and lower bound of each interval are given by the point estimates adjusted by  $z$ times the estimated standard error  (see Equation \ref{eq:ACconf}). 

The standard errors of the generation-effect, the age-effect and the net generation effect  are estimated by 
\begin{equation}\label{eq:sigGE}
\widehat{\sigma}_{\text{GE},\text{gender},\text{late and early gen},t}=\sqrt{\widehat{\sigma}^2_{\text{PS},\text{gender},\text{late gen},t} + \widehat{\sigma}^2_{\text{PS},\text{gender},\text{early gen},t}} \;,
\end{equation} 
\begin{equation}\label{eq:sigAE}
\widehat{\sigma}_{\text{AE},\text{gender},\text{gen},t,\Delta t}=\sqrt{\widehat{\sigma}^2_{\text{PS},\text{gender},\text{gen},t} + \widehat{\sigma}^2_{\text{PS},\text{gender},\text{gen},t+\Delta t} - 2 \rho  \widehat{\sigma}_{\text{PS},\text{gender},\text{gen},t}  \widehat{\sigma}_{\text{PS},\text{gender},\text{gen},t+\Delta t} } \;, 
\end{equation} 
\begin{equation}\label{eq:sigNGE}
\widehat{\sigma}_{\text{NGE},\text{gender},\text{late and early gen},t,\Delta t}=\sqrt{\widehat{\sigma}^2_{\text{GE},\text{gender},\text{late and early gen},t} + (10/7)^2 \widehat{\sigma}^2_{\text{AE},\text{gender},\text{gen},t,\Delta t}} \;,  
\end{equation} 
respectively.

With the above notations (simplified by omitting the time indices), our \textit{hypotheses} about the Boomers can be formalized as     
(i) $H^{mB}_0: {NGE}_{\text{male}, \text{Boomers}}=0$ with the alternative of either  $H^{mB-}_1: {NGE}_{\text{male}, \text{Boomers}}<0$, 
or  $H^{mB+}_1: {NGE}_{\text{male}, \text{Boomers}}>0$; \\ 
(ii) $H^{fB}_0: {NGE}_{\text{female}, \text{Boomers}}=0$ with the alternative  of either  $H^{fB-}_1: {NGE}_{\text{female}, \text{Boomers}}<0$, 
or  $H^{fB+}_1: {NGE}_{\text{female}, \text{Boomers}}>0$. 

Whereas  our second set of hypotheses (about the GenerationX) are \\ 
(iii) $H^{mX}_0: {NGE}_{\text{male}, \text{GenX}}=0$ with the alternative  of  $H^{mX+}_1: {NGE}_{\text{male}, \text{GenX}}>0$; \\ 
(iv) $H^{fX}_0: {NGE}_{\text{female}, \text{GenX}}=0$ with the alternative  of  $H^{fX+}_1: {NGE}_{\text{female}, \text{GenX}}>0$. 

What outcomes of the tests would provide empirical support to the application of the NM and what outcomes would favor the IPF?  
Based on the findings of \cite{NaszodiMendonca2021} presented in Figure \ref{fig:SHC_19802010}, if  $H^j_0$  were rejected for all $j\in\{mB; fB; mX; fX\}$ in favor of  $H^{mB-}_1$, $H^{fB-}_1$,   $H^{mX+}_1$ and $H^{fX+}_1$, then our set of tests  would support the NM. 
Similarly, if $H^{mB}_0$ and $H^{fB}_0$ were rejected in favor of  $H^{mB+}_1$ and $H^{fB+}_1$, respectively, while  $H^{mX}_0$ and   $H^{fX}_0$  were accepted, then 
our tests would support  the IPF. 

There are some more potential outcomes that favor either the NM, or the IPF.  
This is because both the NM and the IPF make predictions on the changes of the equilibrium in the marriage market, rather than on its two determinants, i.e.,  the  men's side of the market and the women's side of the market.   
The sign of the change in the share of homogamous couples under the equilibrium is the same as the sign of change in aggregate preferences at the men's  side of the market provided  the  preferences are unchanged at the women's side of the market. 
E.g., if only men become more (/less) ``picky'' then the  share of homogamous couples increases (/decreases).  
Similarly, if men's aggregate preferences are unchanged from one generation to another, while women's preferences change in a certain way, then the latter determines the direction of change in the equilibrium. 

Accordingly, the potential outcomes of the tests that are also in favor of the NM are the following. 
$H^{mB}_0$ is rejected in favor of the alternative hypothesis of $H^{mB-}_1$, while $H^{fB}_0$ is accepted. 
$H^{fB}_0$ is rejected in favor of the alternative hypothesis of $H^{fB-}_1$, while $H^{mB}_0$ is accepted.   
$H^{mX}_0$ is rejected in favor of the alternative hypothesis of $H^{mX+}_1$, while $H^{fX}_0$ is accepted. 
$H^{fX}_0$ is rejected in favor of the alternative hypothesis of $H^{fX+}_1$, while $H^{mX}_0$ is accepted. 

Similarly, we would get empirical support for the IPF  in the following cases as well:   
$H^{mB}_0$ is rejected in favor of the alternative hypothesis of $H^{mB+}_1$, while $H^{fB}_0$ is accepted;   
$H^{fB}_0$ is rejected in favor of the alternative hypothesis of $H^{fB+}_1$, while $H^{mB}_0$ is accepted.

\begin{table}[ht]
	\caption{The potential and the actual outcomes of the hypothesis tests for the \textit{Boomers}}																							
	\begin{tabular}{llcccc} 
		\hline \hline
		& \multirow{1}{*}{male Boomers } &  & {{$H^{mB-}_1:$}} & $H^{mB}_0:$ & $H^{mB+}_1:$    \\
		&  $\widehat{NGE}=-24.2pp$                              & & {${NGE}<0$} &  ${NGE}=0$ &  ${NGE}>0$    \\ 
		female Boomers&  &  & {$(p_{\rho=0}=0.7\%)$} &     &     \\  
		$\widehat{NGE}=-6.6pp$ 	&  &  & {$(p_{\rho=1}=0.2\%)$}  &     &     \\  \hline
		
		\multicolumn{2}{l}{\multirow{2}{*}{{$H^{fB-}_1$}: ${NGE}<0$}} & $(p_{\rho=0}=22.0\%)$  &   \multirow{2}{*}{NM}  & \multirow{2}{*}{NM}  &  \multirow{2}{*}{}  \\
		&                           &   $(p_{\rho=1}=18.5\%)$   &     &  &         \\

		\multicolumn{2}{l}{\multirow{2}{*}{{$H^{fB}_0\;\;\;\;$}: ${NGE}=0$}} &      & \multirow{2}{*}{{NM}}      &      \multirow{2}{*}{}           & \multirow{2}{*}{IPF}  \\
		&	&                          &       &  &   \\  
		\multicolumn{2}{l}{\multirow{2}{*}{$H^{fB+}_1 \;: {NGE}>0$}} &    & \multirow{2}{*}{}  & \multirow{2}{*}{IPF}  & \multirow{2}{*}{IPF}  \\ 
		&	&                            &    &   &   \\  \hline \hline
	\end{tabular}
	\label{tab:HB}
	\textit{Note}:  this table lists all theoretically possible outcomes of the tests. Some support the application of the NM, while some others support the IPF.

	The p-values report the actual outcomes of the tests under two extreme values of the correlation. The value $\rho=1$ (/$\rho=0$) suggests that there is a perfect (/no) correlation between the empirical shares of the ``picky'' individuals belonging to the same generation while being observed in 2010 and 2017.   
\end{table}

\vspace{1cm}


While the detailed results of our tests are presented in 
 Appendix C, Tables \ref{tab:HB} and \ref{tab:HX} summarize them. 
In particular, Table \ref{tab:HB} shows that at any significance level above 0.7\%, \textit{our tests on the Boomers support the application of the NM},  irrespective of the calibrated value of the correlation $\rho$. This is because the point estimates for the net generation effects are $-24.2$ and $-6.6$ percentage points for the male and the female Boomers, respectively;   while the p-values are in the $\left[0.2\%, 0.7\% \right]$ and $\left[18.5\%, 22\% \right]$  intervals for men and women, respectively. (The exact p-values within their intervals depend on $\rho$).  The tests, -- run separately for men and women --,  accept the hypotheses  $H^{mB-}_1$ and  $H^{fB}_0$ for $0.7\%<\alpha/2<18.5\%$, whereas they accept the hypotheses $H^{mB-}_1$ and $H^{fB-}_1$  for $22\%<\alpha/2$.

Similarly to the Boomers, it is the test on the male survey respondents in the GenerationX that plays the primary role at the method-selection (see Table \ref{tab:HX}).    
At the 20\% significance level, we can reject $H^{mX}_0$ and the IPF in favor of $H^{mX+}_1$ and the NM even under the zero correlation assumption since the related p-value is in the range of [16.1\%, 18.9\%].

{Testing at the 20\% significance level may be viewed to be unusual by researchers caring primarily about the Type I error, e.g., in the context of statistically proving the effectiveness of a pharmacy. However, this choice is reasonable if the test is used for validating a model (or a method) against another model (or a method), while committing the Type I error and the Type II error are perceived to be equally costly. Under the choice of $\alpha/2=20\%$, our test is even conservative since  the probability of the Type I error is 20\%, while the probability of the Type II error is higher than 34\%.} 

\begin{table}[ht]
	\caption{The potential and the actual outcomes of the hypothesis tests for the \textit{GenerationX}}
	\begin{tabular}{llccc} 
		\hline \hline
		& \multirow{1}{*}{male GenX} &  &  {$H^{mX}_0:$}  & {{ $H^{mX+}_1:$}}    \\
		&	 $\widehat{NGE}=9.8pp$                           &  & ${NGE}=0$ &  {${NGE}>0$}    \\  
		female GenX&                   &  &     &    $(p_{\rho=0}=18.9\%)$  \\  
		$\widehat{NGE}=2.7pp$	&                   &  &     &  $(p_{\rho=1}=16.1\%)$    \\  \hline

		\multicolumn{2}{l}{\multirow{2}{*}{{{$H^{fX}_0\;\;\;$}:$\;{NGE}=0$}}}  &  &  \multirow{2}{*}{IPF}   & \multirow{2}{*}{{NM}}    \\
		&	&                            &   &    \\  
		
		\multicolumn{2}{l}{\multirow{2}{*}{{$H^{fX+}_1: {NGE} >0$}}}   &  {$(p_{\rho=0}=40.0\%)$}      & \multirow{2}{*}{NM}    &  \multirow{2}{*}{NM}   \\ 
		&	&  {$(p_{\rho=1}=38.8\%)$}                          &     &   \\    \hline \hline
	\end{tabular}\\
	\label{tab:HX}
	\textit{Note}:  same as under Table \ref{tab:HB}. 
\end{table}

\section{\textbf{Interpreting the results in a broader context}}\label{sec:disc}

As we have seen, our survey data  analysis  facilitates method-selection. 
Method-selection is seemingly a technical task.  
However, it has relevance even for policy-making for the following reasons.

\textit{The disagreement between  the IPF  and the NM over the trend of homophily  
means a disagreement over the trend of a specific, non-monetary dimension of inequality.} 
This is because homophily is an indicator of the perceived width of the social gap between ``them'' and ``us''.  

To recall, the disagreements between the two methods are the following. 
According to the NM and also according to the rich survey data analyzed in this paper, \textit{the studied dimension of inequality displayed a U-shaped pattern} over the second half of the twentieth century and the first decade of the twenty-first century in the US:    
 when the late Boomers gradually replaced the early Boomers on the marriage market, both the revealed and the self-reported marital homophily has been moderated.  
 Later, when the late GenerationX gradually replaced the early GenerationX on the marriage market, both the revealed and the self-reported homophily has been strengthened.    

In contrast to the NM, the \textit{IPF suggests that the trend of the studied dimension of inequality was positive in the first period, while inequality was stagnant in the second period}. 
So, the application of the IPF may lead policymakers to believe that not even the generous welfare policies, from which the American late boomers benefited more than the early boomers,\footnote{For instance,  
	many more young people in the early boomer generation received student benefit than young people in the late boomer generation (see \citealp{Dynarski2003}).}  
could close the social gaps.   
Similarly, the IPF supports the view that the late genXers were as cohesive in 2010 as the early genXers were in 2000, according to their marital sorting. Based on this finding, policymakers may acquire the misconception that not even the skyrocketing economic inequality during the Great Recession could damage social cohesion in the US. Again, \textit{both controversial views gain support provided one applies the IPF rather than the NM}.

It is the \textit{responsibility of the research community to weed out obsolete methods and disproven views that may even be harmful to society}. This paper serves this scientifically and socially desirable goal in a constructive manner: based on our survey-based method-selection, we propose the application of the NM-method instead of the IPF.

	For future policies, the significance of our method-selection is due to the fact that the choice between the IPF and the NM makes a difference not only to what economic and social programmes are believed to have been effective in decreasing inequality in the past and whether the Great Recession following the Global Financial Crisis has increased inequality. It also makes a difference to what future paths are believed to be possible.

As to the historical trend in sorting, there  is a growing agreement in the literature about the pattern of some other, but related dimensions of inequality.\footnote{As it is argued by \cite{NaszodiMendonca2019}, educational assortative mating and economic inequality are closely related. First, one's education level is a proxy for one's ability to generate income and accumulate wealth.  Second, they show that the employment gap between different education groups (i.e., the difference between the education group-specific chances of being employed) is highly correlated with the degree of sorting along the educational characterized by the LL-measure.}  
In particular, there is a forming consensus about the U-curve historical trend in  \textit{wage},  \textit{income} and \textit{wealth  inequality}.  
\cite{goldin2000}, \cite{PikettySaez2003}, and \cite{SaezZucman2016}  
contributed largely to shaping this new consensus.  
They used wage data and American tax records, rather than surveys,  
to estimate the  wage, income and wealth distributions for a period covering also the decades analyzed in this paper.\footnote{We stress that the U-curve pattern itself has not been challenged even though there is an ongoing debate in the literature on how pronounced the decline and the subsequent increase in inequality were (see \citealp{Bricker2016}, \citealp{AutenSplinter2022},  \citealp{Geloso2022}).} 

The discourse in the assortative mating literature lags far behind the debate about the monetary dimensions of inequality since there is no agreement yet over the qualitative trend in  educational homophily and its measurement.  Although the process of forming the related consensus is delayed, it will happen sooner or later.

\section{\textbf{Conclusion}}\label{sec:concl}

Population data on couples     
are informative about a specific dimension of inequality, i.e.,  the intensity of homophily in a society.    
However, to identify historical changes in this dimension of inequality,      
one does not only need  data on marriages and cohabitations, 
but also a method  appropriate  for disentangling ``desires'' and ``opportunities''.         
The commonly used IPF algorithm and its alternative method, the NM, seem to be natural candidates for performing the related counterfactual decomposition.

In this paper, we highlighted the difference between the two sets of problems these methods were original developed for. 
In addition,  we presented some theoretical considerations on the basis of which it is questionable that the IPF is  suitable for constructing counterfactuals in general.    
Then, we presented an empirical method-selection approach.   
Our approach is similar to the one proposed by \cite{NaszodiMendonca2021}. 

Their selection criteria, as well as ours, exploit the fact that the two competing methods disagree on the relative strength of aggregate revealed preferences in some generations whose aggregate self-reported preferences are known from a survey.           
The survey data they use is from a single wave. 
Their survey evidence  seem to corroborate the application of the NM.
However, this evidence is subject to a criticism: 
the variation in the self-reported preferences across different generations identified from only one wave of a survey  
may come partly from the variation of preferences over the course of individuals' lives.       

In this paper, we used survey data from two waves to net the generation-effects from the confounding age-effects. 
In addition, unlike \cite{NaszodiMendonca2021}, we conducted some formal hypothesis tests about the sign and magnitude of the net generation-effects.    
Our tests provide even more convincing evidence in favor of the 
U-shaped historical trend in the specific dimension of inequality studied by the assortative mating literature. 
Thereby, our paper offers even more convincing support for the NM than the survey statistics presented by \cite{NaszodiMendonca2021}.

\bibliography{Bib_IPF_limitations}

\newpage

\begin{appendices}
	
	\renewcommand{\theequation}{A\arabic{equation}}
	\renewcommand{\thetable}{A\arabic{table}}
	\renewcommand{\thefigure}{A\arabic{figure}}

	\setcounter{equation}{0}
	\setcounter{table}{0}
	\setcounter{figure}{0}
	\setcounter{page}{1}

\begin{center}

{\large 		\textbf{Appendix}\\
	of the paper \\
	\textbf{What do surveys say about the trend in inequality and
the applicability of two table-transformation methods?
}
}

\end{center}

\begin{flushleft}
	{\large \textbf{Appendix A: Counterfactuals constructed by the NM-method}}

\end{flushleft}

In this section of the appendix, we illustrate with the numerical examples introduced in the paper that the NM-method commutes with the operation of merging neighboring columns.\footnote{A modified version of the examples with transposed tables illustrate that the NM also commutes with the operation of merging neighboring rows.}   

To recall, in our first numerical example, the assorted traits are dichotomous.  The  $P$ and $Q$ contingency tables are   
$P^{\text{num1}}=$ 
\scalebox{0.7}{$\begin{bmatrix}
	500   & 700 \\
	100   & 700
	\end{bmatrix}$} 
and 
$Q^{\text{num1}}= {\scriptsize \begin{bmatrix}
	500   & 500  \\
	100   & 900
	\end{bmatrix} }$. 

The NM-method constructs the counterfactual with the following formula (see Eq.15 in \citealp{NaszodiMendonca2021}): \\ 
$Q^{\text{NM,}{\text{num1}}}_{1,1}=\frac{ \left[  Q^{\text{num1}}_{1,1} -\text{int}( Q^{\text{num1}}_{1,\cdot} Q^{\text{num1}}_{\cdot,1}/ Q^{\text{num1}}_{\cdot,\cdot}) \right] \left[\text{min}(P^{\text{num1}}_{1,\cdot}, P^{\text{num1}}_{\cdot, 1})  -\text{int}( P^{\text{num1}}_{1,\cdot} P^{\text{num1}}_{\cdot,1}/ P^{\text{num1}}_{\cdot,\cdot})   \right]      }
{\text{min}(Q^{\text{num1}}_{1,\cdot}, Q^{\text{num1}}_{\cdot, 1}) -\text{int}( Q^{\text{num1}}_{1,\cdot} Q^{\text{num1}}_{\cdot,1}/ Q^{\text{num1}}_{\cdot,\cdot}) }+$
\begin{equation}\label{NM} 
 + \text{int}( P^{\text{num1}}_{1,\cdot} P^{\text{num1}}_{\cdot,1}/ P^{\text{num1}}_{\cdot,\cdot})
\end{equation}

By substituting  $Q^{\text{num1}}_{1,1}=500$, $Q^{\text{num1}}_{1,\cdot}=1,000$, $Q^{\text{num1}}_{\cdot,1}=600$,  $Q^{\text{num1}}_{\cdot,\cdot}=2,000$, $P^{\text{num1}}_{1,\cdot}=1,200 $, $P^{\text{num1}}_{\cdot, 1}=600$, $P^{\text{num1}}_{\cdot,\cdot}=2,000$ into the NM-formula of (\ref{NM}), we obtain:  
$Q^{\text{NM,}{\text{num1}}}_{1,1}=520$. The other three cells of the $Q^{\text{NM,}{\text{num1}}}$ matrix can be easily calculated from the row sum and column sum restrictions on the matrix. We get 
$Q^{\text{NM,}{\text{num1}}}=  {\scriptsize \begin{bmatrix}
	520   & 680 \\
	\;80   & 720
	\end{bmatrix} }$. 
This is the counterfactual joint distribution obtained with the NM-method in the first numerical example.   

Now, let us see what counterfactual joint distribution is constructed by NM-method in the second numerical example. 
To recall, husbands' trait  is dichotomous, while wives' trait is trinomial -- taking the values low (L), medium (M) and high (H) --  
in the second example.  
Tables $P$ and $Q$ are given by the following 2-by-3 tables: $P^{\text{num2}} =  
\begin{array}{c}
\text{\scriptsize{L}} \;\; \quad \text{\scriptsize{M}} \quad \;\; \text{\scriptsize{H}} \\
 {\left[\begin{array}{ccc}
500 & 400 & 300 \\
100 & 400 & 300
\end{array}\right]}
\end{array}$ 
and\vspace{-8mm} \\  
$Q^{\text{num2}}= 
\begin{array}{c}
\text{\scriptsize{L}} \;\; \quad \text{\scriptsize{M}} \quad \;\; \text{\scriptsize{H}} \\
\left[\begin{array}{ccc}
	500   & 300 &  200 \\
100   & 300 & 600
\end{array}\right]
\end{array}$.  
To solve this problem with the NM, we have to dichotomize these matrices.  
It can be done in two different ways.  
In the first dichotomization, wives with medium level of education are reclassified to be highly educated. 
The corresponding 2-by-2 matrices are:  
$P^{\text{num2, dich1}}= \begin{array}{c}
\text{\scriptsize{L}} \; \quad \text{\scriptsize{M+H}} \\
\left[\begin{array}{cc}
500 & 700 \\
100 & 700
\end{array}\right]
\end{array}$ 
and 
$Q^{\text{num2, dich1}}= \begin{array}{c}
\text{\scriptsize{L}} \;\; \quad \text{\scriptsize{M+H}} \\
\left[\begin{array}{cc}
500 & 500 \\
100 & 900
\end{array}\right]
\end{array}$. 

In the second dichotomization, wives with medium level of education are reclassified to be low educated. 
The corresponding 2-by-2 matrices are:  
$P^{\text{num2, dich2}}= \begin{array}{c}
\text{\scriptsize{L+M}}  \quad \;\text{\scriptsize{H}} \\
\left[\begin{array}{cc}
900 & 300 \\
500 & 300
\end{array}\right]
\end{array}$ 
and\vspace{-8mm} \\ 
$Q^{\text{num2, dich2}}= \begin{array}{c}
\text{\scriptsize{L+M}}  \quad \;\text{\scriptsize{H}} \\
\left[\begin{array}{cc}
800 & 200 \\
400 & 600
\end{array}\right]
\end{array}$. 

By applying the NM-formula of (\ref{NM}) for the pair of $P^{\text{num2, dich1}}$ and $Q^{\text{num2, dich1}}$,  and the pair  of 
 $P^{\text{num2, dich2}}$ and $Q^{\text{num2, dich2}}$, we obtain 
$Q^{\text{NM,}{\text{num2, dich1}}}=  \begin{array}{c}
\text{\scriptsize{L}}  \quad \text{\scriptsize{M+H}} \\
\left[\begin{array}{cc}
520 & 680 \\
\;80 & 720
\end{array}\right]
\end{array}$ and\vspace{-8mm} \\  
$Q^{\text{NM,}{\text{num2, dich2}}}=  \begin{array}{c}
\text{\scriptsize{L+M}}  \quad \text{\scriptsize{H}} \\
\left[\begin{array}{cc}
1,020 & 180 \\
\;\;380 & 420
\end{array}\right]
\end{array}$, respectively. 

We can calculate the NM counterfactual table in the original 2-by-3 problem from $Q^{\text{NM,}{\text{num2, dich1}}}$ and $Q^{\text{NM,}{\text{num2, dich2}}}$. 
It is 
$Q^{\text{NM,}{\text{num2}}}= \begin{array}{ccc}
\text{\scriptsize{L}} \;\; \quad \text{\scriptsize{M}} \quad \;\; \text{\scriptsize{H}} \\
\left[\begin{array}{ccc}
520 & 500 & 180 \\
80 & 300 & 420
\end{array}\right]
\end{array}$. 

By merging the last two columns of table $ Q^{\text{NM,}{\text{num2}}}$, we obtain: 
$Q^{\text{NM,}{\text{num2}}, \text{merged}}=  \begin{array}{c}
\text{\scriptsize{L}}  \quad \text{\scriptsize{M+H}} \\
\left[\begin{array}{cc}
520 & 680 \\
\;80 & 720
\end{array}\right]
\end{array}$. 
Note, this table is the same as $Q^{\text{NM,}{\text{num1}}}$. So, irrespective of first merging the last two columns of $P^{\text{num2}}$ and $Q^{\text{num2}}$  before applying the NM, or the other way around, we obtain the same counterfactual table.\\



\begin{flushleft}
	{\large \textbf{Appendix B: The distributions of survey participants}}

\end{flushleft}

Let us visit the  \textit{distribution of survey participants} by gender, generation  and also by the dichotomous type of respondent--non-respondent.  
In 2010, the survey question was answered by 289 women and 237 men in the four age groups studied by \cite{NaszodiMendonca2021}. Out of the 289 women, 84 were in the age group 60--64 (representing early Boomers), 92 were in the age group 50--54 (representing late Boomers), 60 were in the age group 40--44 (representing early GenerationX) and 53 were in the age group 30--34 (representing late GenerationX).  Out of the 237 men respondents, 56 were in the age group 60--64, 75 were in the age group 50--54, 61 were in the age group 40--44, 45 were in the age group 30--34. Answering the survey question was refused by only 3 women (of age 35, 55, and 87 years) and 1 men (of age 57 years). 

In 2017, the question was answered by 809 female Boomers and 754 male Boomers (born between 1946--1964);  
715 females in GenerationX  and 756 males in GenerationX (born between 1965--1980). In the same year, answering the survey question was refused by 3 female Boomers,  2 male Boomers,  and by 1 female in GenerationX.  

In 2010, the generational distribution of the survey respondents was this: there were 302 female Boomer respondents and  271 male Boomer respondents (born between 1946--1964);   188 female respondents in GenerationX and 176 males respondents in GenerationX (born between 1965--1980). The generational and gender distribution of the survey non-respondents was the following: 1 female Boomer, 1 male Boomer  and 1  female in GenerationX.\\


\begin{flushleft}
	{\large \textbf{Appendix C: Detailed empirical results}}

\end{flushleft}

The results of our empirical analysis are presented by Tables \ref{tab:distr_boomer}, \ref{tab:distr_boomer_rho1} and \ref{tab:distr_genx}, \ref{tab:distr_genx_rho1} for the Boomers and the GenerationX, respectively. 


Let us first look at the estimates of the \textit{population-shares} of ``picky'' individuals among the \textit{Boomers} (see column 6 in  Table \ref{tab:distr_boomer}). 
Apparently,  this share  was significantly lower  among men in the generation of the late Boomers  in 2010 than among men in the  generation of the early Boomers in the same year.   
For women, the population-share of ``picky'' individuals was not significantly different among the late  Boomers and the early Boomers in 2010  since  the 60\% confidence interval of 
${GE}_{\text{female}, \text{late and early GenX}, 2010}$ 
contains the value zero.

Next, let us visit the \textit{age-effect} of the \textit{Boomers} (see column 7 of Table \ref{tab:distr_boomer}).  
It informs us about how the Boomers' responses have changed over seven years between 2010 and 2017. 
The age-effect is negative both for men and women.\footnote{This finding is line with a by-product of the analysis by \cite{Eika2019}: their assortativity indicator is adjusted by a negative age-effect (see Fig. 4 \citealp{Eika2019}).}  
So, as the Boomers get older, fewer of them tend to report education to be a very important spousal trait.

Accordingly,   
the late Boomer men are found to be even less  ``picky'' about spousal education than the early Boomer men  on average,     
once the age-effect is controlled for  (see column 8 in the upper part of Table \ref{tab:distr_boomer}).  
As to the confidence interval of the \textit{net generation-effect of male Boomers}, its upper bound is decreasing in the correlation $\rho$.  
It is -15.9\% for the unreasonably low value of zero correlation   (see column 8 in the upper part fo Table \ref{tab:distr_boomer}), while it is -16.8\% for the  other extreme of perfect correlation (see Table \ref{tab:distr_boomer_rho1}). 
The p-value 
is  as low as 0.7\% when the correlation is calibrated to zero. 
Whereas it is 0.2\% for the correlation calibrated to one. 
So, irrespective of the  correlation's value,  we can reject $H^{mB}_0$ in favor of  $H^{mB-}_1$  at any meaningful significance level.

For \textit{women}, the adjustment with the age-effect results in a confidence interval of the \textit{net generation-effect} that is entirely in the negative range if the  correlation is one.    
However, it contains the value zero  in the benchmark case of zero correlation  (see column 8 in the lower part fo Table \ref{tab:distr_boomer}). 
The p-value is 18.5\% if the correlation is one, while it is 22\% if the correlation  is zero. 

So, we can reject $H^{fB}_0$ at any significance level above  22\%, irrespective of the calibrated value of  $\rho$.  
As a reference of comparison of the 22\%, we calculate the crossover error rate, i.e.,  
the significance level that equates the size of the test of $H^{fB}_0: {NGE}_{\text{female}, \text{Boomers}}=0$ with its $\beta$ (=1-power) under the alternative of $H^{fB-}_1: {NGE}_{\text{female}, \text{Boomers}}=-6.6$ percentage points. It is higher than 22\% as it is 30\%. 

Finally, it is worth to note that even if we set the bound on the Type I error lower than 22\% (at the cost of increasing the probability of the Type II error above 45\%) and accept  $H^{fB}_0$, our tests for the male Boomers and the female Boomers together clearly support the application of the NM for constructing counterfactuals. This is because revealed homophily should be found to be weaker among the late Boomers relative to the early Boomers even if only the late Boomer males' self-reported preferences  are found to be weaker than that of the early Boomer males, while the self-reported preferences of the 
late and early Boomer females are found similar.

Let us turn to the survey responses of the \textit{GenerationX}.  
Similarly to the age-effects of the Boomers, the age-effects for men  and women are negative in the GenerationX.     
So, as the members of the GenerationX get older, fewer of them tend to report education to be a very important spousal trait (see column 7 in  Table \ref{tab:distr_genx}).    
Therefore, the adjustment with the age-effects works against accepting  $H^{mX+}_1$  and $H^{fX+}_1$. 
For instance, the point estimates of males' generation-effect is   
$\widehat{GE}_{\text{males}, \text{late and early GenX}, 2010}=13.2\%$, whereas their net generation-effect (i.e., the generation-effect adjusted with the age-effect) is lower. The latter is        
$\widehat{NGE}_{\text{males}, \text{late and early GenX}, 2010, 2020}=9.8\%$ (see columns 6 and 8 in the upper part of Table \ref{tab:distr_genx}).

Still, despite the negative sign of the estimated age-effect, the point estimates of males' net generation-effect suggests that the share of ``picky'' men would have been substantially  higher (by almost 10 percentage points) in the late GenerationX relative to the  early GenerationX provided the members of these generations had been observed at the same age. 

As to the confidence interval of ${NGE}_{\text{males}, \text{late and early GenX}, 2010, 2020}$, it  is in the positive range. 
Its lower bound is increasing in $\rho$. 
The lower bound is 0.4\% for the unreasonably low value of zero correlation (see column 8 in the upper part of Table \ref{tab:distr_genx}), while it is  1.5\% for the intuitively more reasonable other extreme of perfect correlation (see Table \ref{tab:distr_genx_rho1}).    
So, we can reject $H^{mX}_0$ at the 20\% significance level  in favor of $H^{mX+}_1$  even under the zero correlation assumption.    
(The p-value is 18.9\% under zero correlation, whereas  it is 16.1\% under perfect correlation. The crossover error rate is around 34\%.)

By contrast, we cannot reject  $H^{fX}_0$  against  $H^{fX+}_1$  for women at any significance level lower than 20\%  since  the 60\% symmetric confidence interval of their net generation-effect contains the value zero   (see column 8 in the lower part of Table \ref{tab:distr_genx}).

To conclude, although women's declared preferences do not seem to have changed as remarkably as men's stated preferences did across the generations studied,\footnote{Interestingly, there is much less inter-generational  variation in social hierarchy  among female baboons  relative to male baboons. ``The ranks of female juveniles depend more on their mothers' ranks, while the ranks of male juveniles depend more on their own size and fighting ability'' (see: \url{https://www.princeton.edu/~baboon/social_life.html}). If humans resemble in this respect these primates and the (non-)persistence of social status is indicative about the (non-)persistence of marital aspiration then it explains why men's preferences vary more from one generation to another than women's  preferences do.}      
we find the late Boomers to be less ``picky'' overall than the early Boomers, while we find the late GenerationX to be much more ``picky'' overall than the early GenerationX 
because of the change in preferences  at the men's side of the marriage market.  
These results validate the NM.

\begin{landscape}																										
	\begin{table}[ht]																									
		
		\caption{The  views of  the opposite sex in the generation of \textit{Boomers} on the {importance} of spousal education}																							
		\begin{tabular}{clccccccccc}																						
			
			\hline	\hline																							
			\multicolumn{1}{l}{}                	&	 Generation 	&	 Survey  	&	 Num.    	&	 Num.    	&	Share of     	&	Estimated	&	 Generation-     	&	 Age-         	&	 Net  \\						
			\multicolumn{1}{l}{}                	&		&	  year   	&	 of   res-     	&	 of   res-     	&	``picky''	&	population-	&	  effect         	&	  effect      	&	  generation-  \\						
			\multicolumn{1}{l}{}                	&		&	         	&	 ponses          	&	 ponses          	&	respon-	&	share**	&	 with  	&	 (older-      	&	   effect \\ 						
			\multicolumn{1}{l}{}                	&		&	         	&		&	``very 	&	dents*	&		&	 age-effect     	&	 younger)     	&	\\						
			\multicolumn{1}{l}{}                	&		&	         	&	           	&	impor-	&		&		&	 (younger-            	&	              	&	 (younger-\\  						
			\multicolumn{1}{l}{}                	&		&	         	&	           	&	tant''	&		&		&	 older)      	&	              	&	  older)\\  						
			\multicolumn{1}{l}{}                	&		&	         	&	           	&		&		&	\textit{[60\% conf. }	&	\textit{[60\% conf. }	&	\textit{[60\% conf. }	&	\textit{[60\% conf. }			\\			
			\multicolumn{1}{l}{}                	&		&	         	&	           	&		&		&	\textit{interval]}	&	\textit{interval]}	&	\textit{interval]}	&	\textit{interval]}			\\			
			\multicolumn{1}{l}{}                	&	 \textit{(Year of birth)}	&		&		&		&	(in \%)	&	(in \%)	&	(in pp)	&	(in pp)	&	(in pp)			\\			
			\multicolumn{1}{l}{}                	&		&	(1)	&	(2)	&	(3)	&	(4)=(3)/(2)	&	(5)	&	(6)	&	(7)	&	  (8)=(6)+(7)$\times$10yrs/7yrs\\   \hline

			\multirow{8}{*}{\rotatebox{90}{Male respondents}} 																								
			
			&	 Late Boomer  	&	2010	&	75	&	25	&	33.3	&	33.5	&	 \multirow{3}{*}{ \Huge{ \}} \Large{-11.2}}                                                    	&	 \multicolumn{1}{l}{}                                  	&	\multirow{3}{*}{   \Huge{-24.2}}                                                 \\						
			&	\textit{ (1956-1960)}     	&		&		&		&		&	 \textit{[28.9;38.1]}	&		&		&						\\	
			&	 Early Boomer 	&	2010	&	56	&	25	&	44.6	&	44.7	&	                                                                           	&	 \multicolumn{1}{l}{}                                  	&	                                                                           \\						
			&	\textit{ (1946-1950)}     	&		&		&		&		&	\textit{[39.2;50.3]}	&	\textit{[-18.4;-4.0]}	&		&	\textit{[-32.5;-15.9]}				\\		
			&	 Boomer            	&	2010	&	271	&	104	&	38.4	&	38.4	&	 \multicolumn{1}{l}{}                                                      	&	\multirow{3}{*}{ \Huge{ \}} \Large{-9.1}}                                                    	&	 \multicolumn{1}{l}{}                                                      \\						
			&	\textit{ (1946-1964)}     	&		&		&		&		&	\textit{[35.9;40.9]}	&		&		&					\\		
			&	 Boomer            	&	2017	&	754	&	221	&	29.3	&	29.3	&	 \multicolumn{1}{l}{}                                                      	&	                                                       	&	 \multicolumn{1}{l}{}                                                      \\  						
			&	\textit{ (1946-1964) }    	&		&		&		&		&	\textit{[27.9;30.7]}	&		&	\textit{[-11.9;-6.2]}	&					\\	\hline	
			\multirow{8}{*}{\rotatebox{90}{Female respondents}} 																								
			
			&	 Late Boomer  	&	2010	&	92	&	32	&	34.8	&	34.9	&	\multirow{3}{*}{\Huge{ \}} \Large{-3.3}}                                                    	&	 \multicolumn{1}{l}{}                                  	&	\multirow{3}{*}{\Huge{-6.6}}                                                        \\						
			&	\textit{ (1956-1960)}     	&		&		&		&		&	\textit{[30.7;39.1]}	&		&		&					\\		
			&	 Early Boomer 	&	2010	&	84	&	32	&	38.1	&	38.2	&	                                                                           	&	 \multicolumn{1}{l}{}                                  	&	                                                                           \\						
			&	\textit{ (1946-1950)}     	&		&		&		&		&	\textit{[33.8;42.6]}	&	\textit{[-9.4;2.8]}	&		&	\textit{[-13.9;0.0]}				\\		
			&	 Boomer            	&	2010	&	302	&	116	&	38.4	&	38.4	&	 \multicolumn{1}{l}{}                                                      	&	\multirow{3}{*}{  \Huge{ \}} \Large{-2.3}}                                                    	&	 \multicolumn{1}{l}{}                                                      \\						
			&	\textit{ (1946-1964)}     	&		&		&		&		&	\textit{[36.1;40.8]}	&		&		&					\\		
			&	 Boomer            	&	2017	&	809	&	292	&	36.1	&	36.1	&	 \multicolumn{1}{l}{}                                                      	&	                                                       	&	 \multicolumn{1}{l}{}                                                      \\ 						
			&	\textit{ (1946-1964) }    	&		&		&		&		&	\textit{[34.7;37.5]}	&		&	\textit{[-5.1;0.0]}	&					\\	\hline	\hline

		\end{tabular}\\																							
		
		{\raggedright \textit{Notes}: *Proportion of \textit{women} among the survey respondents in a given generation who said that it is a \textit{very important quality} of a good \textit{husband/partner} to be well-educated; or,  																							
			proportion of \textit{men}  who said that it is a very important quality of a good \textit{wife/partner} to be well-educated. 																						
			**Calculated by the approximation proposed by \cite{AgrestiCoull1998}. 																						
			The correlations are assumed to be zero between the empirical shares in 2010 and 2017 among those belonging to the same generation. 
			Following \cite{Leamer1978}, we report the 60\% confidence intervals.\\ 
			\textit{Source:}  Changing American Family survey   and 																						
			The American Trends Panel Wave 28 survey  conducted  by the Pew Research Center 																						
			in 2010 and 2017, respectively (see: { \url{https://www.pewsocialtrends.org/dataset/changing-american-family/}}  and \url{https://www.pewsocialtrends.org/dataset/american-trends-panel-wave-28/}).

			\vspace{10cm} \par }																						
		
		\label{tab:distr_boomer}																								
		
	\end{table}				 																					
\end{landscape}									

\newpage 

\begin{landscape}																										
	\begin{table}[ht]																									
		
		\caption{The  views of  the opposite sex in the \textit{GenerationX} on the {importance} of spousal education}																							
		
		\begin{tabular}{clccccccccc}																																																
			\hline	\hline																							
			\multicolumn{1}{l}{}                	&	 Generation 	&	 Survey  	&	 Num.    	&	 Num.    	&	Share of     	&	Estimated	&	 Generation-     	&	 Age-         	&	 Net  \\						
			\multicolumn{1}{l}{}                	&		&	  year   	&	 of   res-     	&	 of   res-     	&	``picky''	&	population-	&	  effect         	&	  effect      	&	  generation-  \\						
			\multicolumn{1}{l}{}                	&		&	         	&	 ponses          	&	 ponses          	&	respon-	&	share**	&	 with  	&	 (older-      	&	   effect \\ 						
			\multicolumn{1}{l}{}                	&		&	         	&		&	``very 	&	dents*	&		&	 age-effect     	&	 younger)     	&	 \\						
			\multicolumn{1}{l}{}                	&		&	         	&	           	&	impor-	&		&		&	 (younger-            	&	              	&	 (younger-\\  						
			\multicolumn{1}{l}{}                	&		&	         	&	           	&	tant''	&		&		&	 older)      	&	              	&	  older)\\  						
			\multicolumn{1}{l}{}                	&		&	         	&	           	&		&		&	\textit{[60\% conf. }	&	\textit{[60\% conf. }	&	\textit{[60\% conf. }	&	\textit{[60\% conf. }			\\			
			\multicolumn{1}{l}{}                	&		&	         	&	           	&		&		&	\textit{interval]}	&	\textit{interval]}	&	\textit{interval]}	&	\textit{interval]}			\\			
			\multicolumn{1}{l}{}                	&	 \textit{(Year of birth)}	&		&		&		&	(in \%)	&	(in \%)	&	(in pp)	&	(in pp)	&	(in pp)			\\			
			\multicolumn{1}{l}{}                	&		&	(1)	&	(2)	&	(3)	&	(4)=(3)/(2)	&	(5)	&	(6)	&	(7)	&	  (8)=(6)+(7)$\times$10yrs/7yrs\\   \hline

			\multirow{8}{*}{\rotatebox{90}{Male respondents}} 																								
			&	 Late GenX  	&	2010	&	45	&	20	&	44.4	&	44.5	&	\multirow{3}{*}{\Huge{ \}} \Large{13.2}}                                                    	&	 \multicolumn{1}{l}{}                                  	&	\multirow{3}{*}{\Huge{9.8}}                                                        \\						
			&	\textit{ (1976-1980) }    	&		&		&		&		&	\textit{[38.3;50.7]}	&		&		&					\\		
			&	 Early GenX 	&	2010	&	61	&	19	&	31.1	&	31.4	&	                                                                           	&	 \multicolumn{1}{l}{}                                  	&	                                                                           \\						
			&	\textit{ (1966-1970) }    	&		&		&		&		&	\textit{[26.4;36.3]}	&	\textit{[5.2;21.1]}	&		&	\textit{[0.4;19.1]}				\\		
			&	 GenX       	&	2010	&	176	&	67	&	38.1	&	38.1	&	 \multicolumn{1}{l}{}                                                      	&	\multirow{3}{*}{ \Huge{ \}} \Large{-2.4}}                                                    	&	 \multicolumn{1}{l}{}                                                      \\						
			&	\textit{ (1965-1980)}     	&		&		&		&		&	\textit{[35;41.2]}	&		&		&					\\		
			&	 GenX       	&	2017	&	756	&	270	&	35.7	&	35.7	&	 \multicolumn{1}{l}{}                                                      	&	                                                       	&	 \multicolumn{1}{l}{}                                                      \\						
			&	\textit{ (1965-1980)}     	&		&		&		&		&	\textit{[34.3;37.2]}	&		&	\textit{[-5.8; 0.0]}	&					\\	\hline

			\multirow{8}{*}{\rotatebox{90}{Female respondents}} 																								
			&	 Late GenX  	&	2010	&	53	&	24	&	45.3	&	45.3	&	\multirow{3}{*}{\Huge{ \}} \Large{ 5.2}}                                                    	&	 \multicolumn{1}{l}{}                                  	&	\multirow{3}{*}{\Huge{2.7}}                                                        \\						
			&	\textit{ (1976-1980) }    	&		&		&		&		&	\textit{[39.6;51.1]}	&		&		&					\\		
			&	 Early GenX 	&	2010	&	60	&	24	&	40.0	&	40.1	&	                                                                           	&	 \multicolumn{1}{l}{}                                  	&	                                                                           \\						
			&	\textit{ (1966-1970) }    	&		&		&		&		&	\textit{[34.8;45.4]}	&	\textit{[-2.6;13.0]}	&		&	\textit{[-6.5;11.8]}				\\		
			&	 GenX       	&	2010	&	188	&	78	&	41.5	&	41.5	&	 \multicolumn{1}{l}{}                                                      	&	\multirow{3}{*}{ \Huge{ \}} \Large{-1.8}}                                                    	&	 \multicolumn{1}{l}{}                                                      \\						
			&	\textit{ (1965-1980)}     	&		&		&		&		&	\textit{[38.5;44.5]}	&		&		&					\\		
			&	 GenX       	&	2017	&	715	&	284	&	39.7	&	39.7	&	 \multicolumn{1}{l}{}                                                      	&	                                                       	&	 \multicolumn{1}{l}{}                                                      \\						
			&	\textit{ (1965-1980)}     	&		&		&		&		&	\textit{[38.2;41.3]}	&		&	\textit{[-5.2;1.6]}	&					\\	\hline	 \hline

		\end{tabular}\\																																																		
		
		{\raggedright \textit{Notes}: same as under Table \ref{tab:distr_boomer}.																												\vspace{10cm} \par }	
		\label{tab:distr_genx}

	\end{table}				 																						
\end{landscape}																										

\newpage

\begin{landscape}																										
	\begin{table}[ht]																									
		
		\caption{The  views of  the opposite sex in the generation of \textit{Boomers} on the {importance} of spousal education -- \textit{perfect correlation}}																							
		\begin{tabular}{clccccccccc}																						
			
			\hline	\hline																							
			\multicolumn{1}{l}{}                	&	 Generation 	&	 Survey  	&	 Num.    	&	 Num.    	&	Share of     	&	Estimated	&	 Generation-     	&	 Age-         	&	 Net  \\						
			\multicolumn{1}{l}{}                	&		&	  year   	&	 of   res-     	&	 of   res-     	&	``picky''	&	population	&	  effect         	&	  effect      	&	  generation-  \\						
			\multicolumn{1}{l}{}                	&		&	         	&	 ponses          	&	 ponses          	&	respon-	&	share**	&	 with  	&	 (older-      	&	   effect \\ 						
			\multicolumn{1}{l}{}                	&		&	         	&		&	``very 	&	dents*	&		&	 age-effect     	&	 younger)     	&	\\						
			\multicolumn{1}{l}{}                	&		&	         	&	           	&	impor-	&		&		&	 (younger-            	&	              	&	 (younger-\\  						
			\multicolumn{1}{l}{}                	&		&	         	&	           	&	tant''	&		&		&	 older)      	&	              	&	  older)\\  						
			\multicolumn{1}{l}{}                	&		&	         	&	           	&		&		&	\textit{[60\% conf. }	&	\textit{[60\% conf. }	&	\textit{[60\% conf. }	&	\textit{[60\% conf. }			\\			
			\multicolumn{1}{l}{}                	&		&	         	&	           	&		&		&	\textit{interval]}	&	\textit{interval]}	&	\textit{interval]}	&	\textit{interval]}			\\			
			\multicolumn{1}{l}{}                	&	 \textit{(Year of birth)}	&		&		&		&	(in \%)	&	(in \%)	&	(in pp)	&	(in pp)	&	(in pp)			\\			
			\multicolumn{1}{l}{}                	&		&	(1)	&	(2)	&	(3)	&	(4)=(3)/(2)	&	(5)	&	(6)	&	(7)	&	  (8)=(6)+(7)$\times$10yrs/7yrs\\   \hline

			\multirow{8}{*}{\rotatebox{90}{Male respondents}} 																								
			
			&	 Late Boomer  	&	2010	&	75	&	25	&	33.3	&	33.5	&	 \multirow{3}{*}{ \Huge{ \}} \Large{-11.2}}                                                    	&	 \multicolumn{1}{l}{}                                  	&	\multirow{3}{*}{   \Huge{-24.2}}                                                 \\						
			&	\textit{ (1956-1960)}     	&		&		&		&		&	 \textit{[28.9;38.1]}	&		&		&						\\	
			&	 Early Boomer 	&	2010	&	56	&	25	&	44.6	&	44.7	&	                                                                           	&	 \multicolumn{1}{l}{}                                  	&	                                                                           \\						
			&	\textit{ (1946-1950)}     	&		&		&		&		&	\textit{[39.2;50.3]}	&	\textit{[-18.4;-4.0]}	&		&	\textit{[-31.5;-16.8]}				\\		
			&	 Boomer            	&	2010	&	271	&	104	&	38.4	&	38.4	&	 \multicolumn{1}{l}{}                                                      	&	\multirow{3}{*}{ \Huge{ \}} \Large{-9.1}}                                                    	&	 \multicolumn{1}{l}{}                                                      \\						
			&	\textit{ (1946-1964)}     	&		&		&		&		&	\textit{[35.9;40.9]}	&		&		&					\\		
			&	 Boomer            	&	2017	&	754	&	221	&	29.3	&	29.3	&	 \multicolumn{1}{l}{}                                                      	&	                                                       	&	 \multicolumn{1}{l}{}                                                      \\  						
			&	\textit{ (1946-1964) }    	&		&		&		&		&	\textit{[27.9;30.7]}	&		&	\textit{[-10.2;-8]}	&					\\	\hline	
			\multirow{8}{*}{\rotatebox{90}{Female respondents}} 																								
			
			&	 Late Boomer  	&	2010	&	92	&	32	&	34.8	&	34.9	&	\multirow{3}{*}{\Huge{ \}} \Large{-3.3}}                                                    	&	 \multicolumn{1}{l}{}                                  	&	\multirow{3}{*}{\Huge{-6.6}}                                                        \\						
			&	\textit{ (1956-1960)}     	&		&		&		&		&	\textit{[30.7;39.1]}	&		&		&					\\		
			&	 Early Boomer 	&	2010	&	84	&	32	&	38.1	&	38.2	&	                                                                           	&	 \multicolumn{1}{l}{}                                  	&	                                                                           \\						
			&	\textit{ (1946-1950)}     	&		&		&		&		&	\textit{[33.8;42.6]}	&	\textit{[-9.4;2.8]}	&		&	\textit{[-12.9;-0.4]}				\\		
			&	 Boomer            	&	2010	&	302	&	116	&	38.4	&	38.4	&	 \multicolumn{1}{l}{}                                                      	&	\multirow{3}{*}{  \Huge{ \}} \Large{-2.3}}                                                    	&	 \multicolumn{1}{l}{}                                                      \\						
			&	\textit{ (1946-1964)}     	&		&		&		&		&	\textit{[36.1;40.8]}	&		&		&					\\		
			&	 Boomer            	&	2017	&	809	&	292	&	36.1	&	36.1	&	 \multicolumn{1}{l}{}                                                      	&	                                                       	&	 \multicolumn{1}{l}{}                                                      \\ 						
			&	\textit{ (1946-1964) }    	&		&		&		&		&	\textit{[34.7;37.5]}	&		&	\textit{[-3.3;-1.4]}	&					\\	\hline	\hline

		\end{tabular}\\																							
		{\raggedright \textit{Notes}: same as under Table \ref{tab:distr_boomer} except that parameter $\rho$ (the correlation between the empirical shares of the ``picky'' individuals belonging to the same generation while being observed in 2010 and 2017)  is calibrated to one.}\\

		
		\label{tab:distr_boomer_rho1}

	\end{table}				 																					
\end{landscape}

\begin{landscape}																										
	\begin{table}[ht]																									
		
		\caption{The  views of  the opposite sex in the \textit{GenerationX} on the {importance} of spousal education -- \textit{perfect correlation}}																							
		
		\begin{tabular}{clccccccccc}																																																
			\hline	\hline																							
			\multicolumn{1}{l}{}                	&	 Generation 	&	 Survey  	&	 Num.    	&	 Num.    	&	Share of     	&	Estimated	&	 Generation-     	&	 Age-         	&	 Net  \\						
			\multicolumn{1}{l}{}                	&		&	  year   	&	 of   res-     	&	 of   res-     	&	``picky''	&	population	&	  effect         	&	  effect      	&	  generation-  \\						
			\multicolumn{1}{l}{}                	&		&	         	&	 ponses          	&	 ponses          	&	respon-	&	share**	&	 with  	&	 (older-      	&	   effect \\ 						
			\multicolumn{1}{l}{}                	&		&	         	&		&	``very 	&	dents*	&		&	 age-effect     	&	 younger)     	&	 \\						
			\multicolumn{1}{l}{}                	&		&	         	&	           	&	impor-	&		&		&	 (younger-            	&	              	&	 (younger-\\  						
			\multicolumn{1}{l}{}                	&		&	         	&	           	&	tant''	&		&		&	 older)      	&	              	&	  older)\\  						
			\multicolumn{1}{l}{}                	&		&	         	&	           	&		&		&	\textit{[60\% conf. }	&	\textit{[60\% conf. }	&	\textit{[60\% conf. }	&	\textit{[60\% conf. }			\\			
			\multicolumn{1}{l}{}                	&		&	         	&	           	&		&		&	\textit{interval]}	&	\textit{interval]}	&	\textit{interval]}	&	\textit{interval]}			\\			
			\multicolumn{1}{l}{}                	&	 \textit{(Year of birth)}	&		&		&		&	(in \%)	&	(in \%)	&	(in pp)	&	(in pp)	&	(in pp)			\\			
			\multicolumn{1}{l}{}                	&		&	(1)	&	(2)	&	(3)	&	(4)=(3)/(2)	&	(5)	&	(6)	&	(7)	&	  (8)=(6)+(7)$\times$10yrs/7yrs\\   \hline

			\multirow{8}{*}{\rotatebox{90}{Male respondents}} 																								
			&	 Late GenX  	&	2010	&	45	&	20	&	44.4	&	44.5	&	\multirow{3}{*}{\Huge{ \}} \Large{13.2}}                                                    	&	 \multicolumn{1}{l}{}                                  	&	\multirow{3}{*}{\Huge{9.8}}                                                        \\						
			&	\textit{ (1976-1980) }    	&		&		&		&		&	\textit{[38.3;50.7]}	&		&		&					\\		
			&	 Early GenX 	&	2010	&	61	&	19	&	31.1	&	31.4	&	                                                                           	&	 \multicolumn{1}{l}{}                                  	&	                                                                           \\						
			&	\textit{ (1966-1970) }    	&		&		&		&		&	\textit{[26.4;36.3]}	&	\textit{[5.2;21.1]}	&		&	\textit{[1.5;18]}				\\		
			&	 GenX       	&	2010	&	176	&	67	&	38.1	&	38.1	&	 \multicolumn{1}{l}{}                                                      	&	\multirow{3}{*}{ \Huge{ \}} \Large{-2.4}}                                                    	&	 \multicolumn{1}{l}{}                                                      \\						
			&	\textit{ (1965-1980)}     	&		&		&		&		&	\textit{[35;41.2]}	&		&		&					\\		
			&	 GenX       	&	2017	&	756	&	270	&	35.7	&	35.7	&	 \multicolumn{1}{l}{}                                                      	&	                                                       	&	 \multicolumn{1}{l}{}                                                      \\						
			&	\textit{ (1965-1980)}     	&		&		&		&		&	\textit{[34.3;37.2]}	&		&	\textit{[-4; -0.8]}	&					\\	\hline

			\multirow{8}{*}{\rotatebox{90}{Female respondents}} 																								
			&	 Late GenX  	&	2010	&	53	&	24	&	45.3	&	45.3	&	\multirow{3}{*}{\Huge{ \}} \Large{ 5.2}}                                                    	&	 \multicolumn{1}{l}{}                                  	&	\multirow{3}{*}{\Huge{2.7}}                                                        \\						
			&	\textit{ (1976-1980) }    	&		&		&		&		&	\textit{[39.6;51.1]}	&		&		&					\\		
			&	 Early GenX 	&	2010	&	60	&	24	&	40.0	&	40.1	&	                                                                           	&	 \multicolumn{1}{l}{}                                  	&	                                                                           \\						
			&	\textit{ (1966-1970) }    	&		&		&		&		&	\textit{[34.8;45.4]}	&	\textit{[-2.6;13.0]}	&		&	\textit{[-5.4;10.7]}				\\		
			&	 GenX       	&	2010	&	188	&	78	&	41.5	&	41.5	&	 \multicolumn{1}{l}{}                                                      	&	\multirow{3}{*}{ \Huge{ \}} \Large{-1.8}}                                                    	&	 \multicolumn{1}{l}{}                                                      \\						
			&	\textit{ (1965-1980)}     	&		&		&		&		&	\textit{[38.5;44.5]}	&		&		&					\\		
			&	 GenX       	&	2017	&	715	&	284	&	39.7	&	39.7	&	 \multicolumn{1}{l}{}                                                      	&	                                                       	&	 \multicolumn{1}{l}{}                                                      \\						
			&	\textit{ (1965-1980)}     	&		&		&		&		&	\textit{[38.2;41.3]}	&		&	\textit{[-3.3;-0.3]}	&					\\	\hline	 \hline

		\end{tabular}\\																																																		
		{\raggedright \textit{Notes}: same as under Table \ref{tab:distr_boomer_rho1}.}\\


		\label{tab:distr_genx_rho1}

	\end{table}				 																						
\end{landscape}



\end{appendices}

\end{document}